\documentclass[12pt]{article}
\usepackage{axodraw,epsfig}
\usepackage{graphics}
\usepackage{color}

\newcommand{\bea}{\begin{eqnarray}}
\newcommand{\nn}{\nonumber}
\newcommand{\eea}{\end{eqnarray}}
\newcommand{\bd}{\begin{displaymath}}
\newcommand{\ed}{\end{displaymath}}
\newcommand{\be}{\begin{equation}}
\newcommand{\ee}{\end{equation}}
\newcommand{\bi}{\begin{itemize}}
\newcommand{\ei}{\end{itemize}}

\newcommand{\ns}{n\hspace{-1.2ex}/}

\newcommand{\Ds}{D\hspace{-1.5ex}/}

\newcommand{\bfm}[1]{\mbox{\boldmath$#1$}}

\catcode`@=11
\newcount\@tempcntc
\def\@citex[#1]#2{\if@filesw\immediate\write\@auxout{\string\citation{#2}}\fi
  \@tempcnta\z@\@tempcntb\m@ne\def\@citea{}\@cite{\@for\@citeb:=#2\do
    {\@ifundefined
       {b@\@citeb}{\@citeo\@tempcntb\m@ne\@citea\def\@citea{,}{\bf ?}\@warning
       {Citation `\@citeb' on page \thepage \space undefined}}%
    {\setbox\z@\hbox{\global\@tempcntc0\csname b@\@citeb\endcsname\relax}%
     \ifnum\@tempcntc=\z@ \@citeo\@tempcntb\m@ne
       \@citea\def\@citea{,}\hbox{\csname b@\@citeb\endcsname}%
     \else
      \advance\@tempcntb\@ne
      \ifnum\@tempcntb=\@tempcntc
      \else\advance\@tempcntb\m@ne\@citeo
      \@tempcnta\@tempcntc\@tempcntb\@tempcntc\fi\fi}}\@citeo}{#1}}
\def\@citeo{\ifnum\@tempcnta>\@tempcntb\else\@citea\def\@citea{,}%
  \ifnum\@tempcnta=\@tempcntb\the\@tempcnta\else
   {\advance\@tempcnta\@ne\ifnum\@tempcnta=\@tempcntb \else \def\@citea{--}\fi
    \advance\@tempcnta\m@ne\the\@tempcnta\@citea\the\@tempcntb}\fi\fi}
\catcode`@=12

\begin{document}

\begin{titlepage}

\begin{flushright}
\normalsize
PITHA~08/26\\
SFB/CPP-08-81\\
%0810.xxxx [hep-ph]\\
October 18, 2008
\end{flushright}

\vskip1.5cm
\begin{center}
\Large\bf\boldmath
$B\to \chi_{cJ} K$ decays revisited
\unboldmath
\end{center}

\vspace*{0.8cm}
\begin{center}

{\sc M. Beneke}$^{a,b}$ and {\sc L. Vernazza}$^{a}$\\[5mm]
{\small $^a$ \it Institut f{\"u}r Theoretische Physik~E,
    RWTH Aachen University,}\\
{\small\it D--52056 Aachen, Germany}\\[0.2cm]
{\small $^b$ \it 
Institut f\"ur Theoretische Physik, 
Universit\"at Z\"urich,\\
CH -- 8057 Z\"urich, Switzerland}
\end{center}

\vspace*{0.8cm}
\begin{abstract}
  \noindent
  We demonstrate that exclusive $B$ decays to $P$-wave charmonium
  factorize in the non-relativistic limit provided that colour-octet
  contributions are taken into account, and estimate the branching
  fractions. Although there are very large uncertainties, we find 
  reasonable parameter choices, where the main features of the data --  
  large corrections to (naive) factorization and suppression of 
  the $\chi_{c2}$ and $h_c$ final states --  are reproduced though the 
  suppression of $\chi_{c2}$ is not as strong as seen in the data. Our 
  results also provide an example, where an endpoint
  divergence in hard spectator-scattering factorizes and is
  absorbed into colour-octet operator matrix elements.
\vspace*{0.8cm}
%\noindent
%PACS numbers: 12.38.Bx, 14.40.Gx, 14.65.Fy, 14.65.Ha

\end{abstract}

\vfil
\end{titlepage}

\newpage

\section{Introduction}

Exclusive two-body $B$ meson decays $B\to M_1 M_2$ factorize
in the heavy quark limit when the final state meson $M_2$,
that does not pick up the spectator antiquark from the $B$ meson,
is light \cite{Beneke:1999br,Beneke:2000ry,Bauer:2001cu}.
The reason for this is that the quark-antiquark progenitor of
the ``emitted'' meson escapes the decay region as an
object of small transverse size, which remains invisible to
soft-gluon interactions with the $B\to M_1$
transition \cite{Bjorken:1988kk}. This argument extends
to any colour-singlet object $M_2$ that is small compared to
the inverse strong interaction scale, $1/\Lambda$. In
particular, it has been suggested \cite{Beneke:2000ry} that
exclusive decays to charmonia $H$, such as $B\to H K$,
should factorize in the heavy-quark and non-relativistic
limit, when the charmonium radius $1/(m_c v)\ll 1/\Lambda$.
This expectation has been confirmed for decays to the $S$-wave
charmonia $J/\psi$ and $\eta_c$ by explicit next-to-leading
order (NLO)
calculations~\cite{Chay:2000xn,Cheng:2000kt,Song:2002gw}.\footnote{Some
of these papers use a light-cone rather than non-relativistic description
of the charmonium, but this does not affect the conclusion.}
However, when the formalism was applied to $P$-wave charmonium
states~\cite{Song:2002mh,Song:2003yc,Pham:2005ih,Meng:2005fc,Meng:2005er,Meng:2006mi}
infrared (IR) divergences appeared that seem to violate factorization.

In this paper we revisit this problem and show that factorization
is recovered, if one includes the charmonium bound-state scales
$m_c v$, $m_c v^2$ (with $v$ a characteristic velocity of the charm
quark in the bound state, $v\ll 1$) into the theoretical
framework. These scales are assumed to be
intermediate between the heavy quark
masses $m_b$, $m_c$ and the strong interaction scale $\Lambda$.
The divergence structure described above bears resemblance with inclusive
charmonium decay or production in the colour-singlet model.
As is well-known the IR divergence problem in the $P$-wave colour-singlet
amplitudes is resolved by the introduction of colour-octet
operators \cite{Bodwin:1992ye}. We shall see below that a similar
mechanism is at work for $B\to H K$. However, there is an important
difference between $P$-wave charmonium production in inclusive
$B$ decay~\cite{Bodwin:1992qr,Beneke:1998ks} and exclusive decays.
While in the former the charmonium decouples from the inclusive
$B\to X$ transition below the heavy quark mass scale, or at least is
assumed to in previous treatments, this
decoupling takes place for exclusive decays only below the
scale of the binding energy $m_c v^2$, since gluons with this
energy can reconnect to the $BK$ system. In the (formal) heavy-quark
limit this effect is perturbatively calculable since
$m_c v^2\gg \Lambda$, and we shall provide numerical estimates for
this contribution, which has been neglected in previous calculations.
In the real world, $m_c v^2 \sim \Lambda$, and hence reliable
calculations appear to be hard to come by for decays to charmonium,
contrary to decays to light $M_2$.

In Table~\ref{tab1} we collect the current branching fraction measurements
for the decays in question
\cite{Abe:1995aw,Abe:2001mw,Acosta:2002pw,Fang:2002gi,Abe:2002rc,Aubert:2003vc,Aubert:2004gc,Aubert:2004rz,Aubert:2005vwa,Aubert:2005ce,Aubert:2005gw,Soni:2005fw,Aubert:2005vi,Aubert:2006nu,Fang:2006bz,Aubert:2006aj,Aubert:2007qea},
including the $S$-wave final states. By
comparing $\chi_{c1}$ to the $S$-waves, we conclude that the
$P$-wave suppression is almost absent. Recalling that in naive
factorization only the $\chi_{c1}$ state is produced, the pattern of
the $P$-wave results is even more striking, and suggests significant
decay amplitudes beyond naive factorization. The colour-octet
contributions that we identify here may well be the dominant decay
amplitudes, although they turn out to be hard to calculate for
real-world charmonium. Nevertheless, it is interesting to see whether
something can be said from theory that
may help to understand the pattern of experimental data.

\begin{table}
\begin{center}
\begin{tabular}{l|c|c|c|c|c|c} \hline
& $J/\psi$ & $\eta_c$ & $\chi_{c0}$ & $\chi_{c1}$ & $\chi_{c2}$ & $h_c$
\\[0.2cm]
$B^-$ &
$10.26\pm 0.37$ &
$9.8\pm 1.3$ &
$1.88\pm0.30$ &
$5.01\pm0.37$ &
$ < 0.30 $ &
$< 0.38$
\\[0.2cm]
$B_d^0$ &
$\phantom{1}8.63\pm 0.35$ &
$8.7\pm 1.9$ &
$<12.4$ &
$4.0 \pm 0.4$ &
$< 0.41$ &
---
\\[0.2cm]
\hline
\end{tabular}
\caption{Averages of $B\to \mbox{charmonium } + K$ branching fraction
data in units of $10^{-4}$, taken from the Heavy Flavour Averaging 
Group~\cite{HFAG}. The original experimental results can be found in
\cite{Abe:1995aw,Abe:2001mw,Acosta:2002pw,Fang:2002gi,Abe:2002rc,Aubert:2003vc,Aubert:2004gc,Aubert:2004rz,Aubert:2005vwa,Aubert:2005ce,Aubert:2005gw,Soni:2005fw,Aubert:2005vi,Aubert:2006nu,Fang:2006bz,Aubert:2006aj,Aubert:2007qea}.
The limit for $h_c$ assumes $\mbox{Br}\,(h_c\to\eta_c\gamma)=0.5$.
\label{tab1}}
\end{center}
\end{table}

The infrared divergence in hard spectator-scattering is actually an
endpoint divergence in a momentum-fraction convolution integral.
Such endpoint divergences prohibit hard-scattering factorization of
power-suppressed effects in non-leptonic $B$ decays to
light mesons \cite{Beneke:1999br}, and of $B$ to light meson
form factors even at leading power
\cite{Szczepaniak:1990dt,Beneke:2000wa}. Understanding whether and
how such endpoint divergences can be factorized remains a major challenge
to theory. In $B$ decays to $P$-wave charmonia the endpoint
singularity also arises at leading order in the $\Lambda/m_b$ expansion,
but we shall show that it can be factorized into the matrix
elements of colour-octet operators. This is of some conceptual interest, 
since it is not known in general how to factorize endpoint divergences.

Two recent papers \cite{Bobeth:2007sh,Bodwin:2008nf} also discuss
factorization of $B$ decays to charmonium. These papers deal with 
the leading order in the non-relativistic velocity expansion
applicable to decays to $S$-wave charmonia, but do not address
$P$-waves. The term ``factorization'' there refers to the heavy-quark mass
scale $m_b$ or $m_c$, and should thus be distinguished from its
use here. While it is evident that
$B$ decays to $P$-wave charmonia do not factorize at the
heavy-quark mass scale due to the infrared divergences mentioned
above, our concern is to show that perturbative factorization is recovered
when $m_c v^2 \gg \Lambda$ as conjectured in \cite{Beneke:2000ry}.
Corrections to naive factorization for $B$ decays to charmonium have also 
been estimated with light-cone QCD sum 
rules~\cite{Melic:2003bw,Wang:2003gaa,Melic:2004ud}, but with this 
method the issue of infrared singularities in the QCD factorization 
result is not addressed.

\section{Operator definitions and tree-level results}\label{sec2}

\subsection{Effective Hamiltonian and kinematics}
\label{sec2.1}

The effective weak--interaction Hamiltonian for the $b\to c\bar c s$
transition is
\bea
\label{eq:1}
\mathcal{H}_{\mbox{\scriptsize{eff}}}&=&\frac{G_F}{\sqrt{2}}
V_{cb}V^*_{cs} (C_1 Q_1+C_2 Q_2)
\eea
with
\bea
\label{eq:2}
Q_1= (\bar{c}_i b_i)_{V-A}(\bar{s}_j c_j)_{V-A}, \quad
Q_2= (\bar{c}_i b_j)_{V-A}(\bar{s}_j c_i)_{V-A},
\eea
and neglecting the small contributions from the penguin operators.

The following notation is adopted for the kinematics of the two-body
decay process $B\to (\chi_{cJ},h_c) K$:
$p_B$ is the momentum of the $B$ meson, $w^\mu\equiv p_B^\mu/M_B$
with $w^2=1$;
$q$ of the charmonium, with $v^{\mu} \equiv q^{\mu}/M_H$ the
charmonium 4-velocity ($M_H$ being the charmonium mass) with
$v^2=1$; $l^{\mu}$
defines the relative momentum of the $c$ quark inside the charmonium,
so that $p_c=q/2 +l$, $\bar{p}_c=q/2-l$ with $q\cdot l=0$; $p$ is
the momentum of the kaon. Since kaon mass effects can be neglected
in the heavy quark limit, the vector  $n_-^{\mu} =
p^{\mu}/E_K$ ($E_K$ being the kaon energy in the $B$ rest frame)
is light-like. The opposite-pointing light-like
vector is denoted $n_+^\mu$ with $n_-\cdot n_+=2$.
We also define $z=4 m_c^2/m_b^2$, equal to $M_H^2/M_B^2$
up to corrections of order\footnote{Note the dual use of $v$: in
  power-counting estimates $v$ denotes the small {\em relative}
  velocity of the heavy quarks in the charmonium in the  
  rest frame of the charmonium. In kinematical
  relations $v$ is the charmonium velocity vector.} $\Lambda/m_b$ and~$v^2$.
From energy-momentum
conservation
\begin{eqnarray}
\label{kinematics}
&& E_K = \frac{n_+\cdot p}{2} = \frac{M_B^2-M_H^2}{2 M_B},
\qquad n_-\cdot v = \frac{1}{n_+\cdot v} =\frac{M_B}{M_H}.
\end{eqnarray}
For any vector $a$ (or Lorentz index $\mu$) the components
transverse to $n_-$ and $n_+$ are denoted by $a_\perp$ ($\mu_\perp$),
for those orthogonal to $v$ we use $a_\top$ ($\mu_\top$).
Thus
\begin{equation}
a^\mu=a^\mu_\perp + \frac{n_+\cdot a}{2}\,n_-^\mu +
\frac{n_-\cdot a}{2}\,n_+^\mu = a_\top^\mu + v\cdot a\,v^\mu.
\end{equation}

\subsection{SCET/NRQCD operator definitions}

The bottom and charm quark masses are assumed to be heavy, 
with $m_c/m_b$ fixed in the heavy-quark limit. Integrating out
the heavy quark mass scales $m_b$, $m_c$ 
leads to an effective theory, in which the $b$
quark is static as in heavy-quark effective theory, the charm quarks
are non-relativistic (in their center-of-mass frame) as in NRQCD and the
light quarks are collinear (or soft) as in soft-collinear effective theory
(SCET). The situation is similar to the corresponding one for
charmless $B$~decays~\cite{Chay:2003ju,Bauer:2004tj,Beneke:2005vv}, except
that the meson that does not absorb the spectator quark is now described
by non-relativistic rather than collinear fields.

To build the
decay operators in the effective theory below the heavy quark scale,
we introduce the non-relativistic bilinears
\bea
\label{eq:3}
\nn
\mathcal{O}(^1P_1^{(1)})&\equiv&
\psi_v^{\dagger} \left[n_{-\mu}\left(-\frac{i}{2}\right)
\stackrel{\leftrightarrow}{D}_{\top}^{\mu}\gamma_5\right] \chi_v, \\
\nn
\mathcal{O}(^3P_0^{(1)})&\equiv&
\psi_v^{\dagger} \left[-\frac{1}{\sqrt{3}}\left(-\frac{i}{2}\right)
\stackrel{\leftrightarrow}{\Ds}_{\top}\right] \chi_v, \\
\nn
\mathcal{O}(^3P_1^{(1)})&\equiv&
\psi_v^{\dagger} \left[\frac{n_{-\mu}}{2\sqrt{2}} \left(-\frac{i}{2}\right)
\left[\stackrel{\leftrightarrow}{\Ds}_{\top},\gamma_{\top}^{\mu}
\right]\gamma_5\right] \chi_v, \\
\mathcal{O}(^3P_2^{(1)})&\equiv&
\psi_v^{\dagger} \left[n_{-\mu}n_{-\nu}\left(-\frac{i}{2}\right)
\stackrel{\leftrightarrow}{D}_{\top}^{(\mu} \gamma_{\top}^{\nu)}\right] \chi_v,
\nn \\[0.1cm]
\nn
\mathcal{O}(^1S_0^{(8)})&\equiv& \psi_v^{\dagger}
\left[\gamma_5 T^A\right]\chi_v,  \\[0.2cm] \nn
\mathcal{O}(^3S_1^{(8)})&\equiv&\psi_v^{\dagger}
\left[n_{-\mu} \gamma_{\top}^{\mu} T^A \right] \chi_v \\[0.2cm]
\nn
\mathcal{O}^{\mu}(^1S_0^{(8)})&\equiv& \psi_v^{\dagger}
\left[v^{\mu}\gamma_5 T^A\right]\chi_v,  \\[0.2cm]
\mathcal{O}^{\mu}(^3S_1^{(8)})&\equiv&\psi_v^{\dagger}
\left[\gamma_{\top}^{\mu} T^A \right] \chi_v
\eea
of non-relativistic quark ($\psi$) and anti-quark ($\chi$)
spinor fields.\footnote{$\stackrel{\leftrightarrow}{D}_{\top}^{(\mu} 
\gamma_{\top}^{\nu)}$ denotes the symmetric, traceless part of 
the tensor $\stackrel{\leftrightarrow}{D}_{\top}^{\mu} 
\gamma_{\top}^{\nu}$.} It is
convenient to use a covariant generalization of the NRQCD Lagrangian,
where the non-relativistic fields are four-component spinors
satisfying $\not \!v\,\psi_v=\psi_v$ and  $\not \!v\,\chi_v=-\chi_v$.
In the charmonium rest frame $v=(1,\bfm{0})$ the non-zero
spinor components reduce to the familiar non-relativistic two-spinors,
and the $\top$ components of a contravariant index equal the
spatial components.

\begin{figure}[t]
  \begin{center}
  \vspace*{-1.3cm}
  \includegraphics[width=1.2\textwidth]{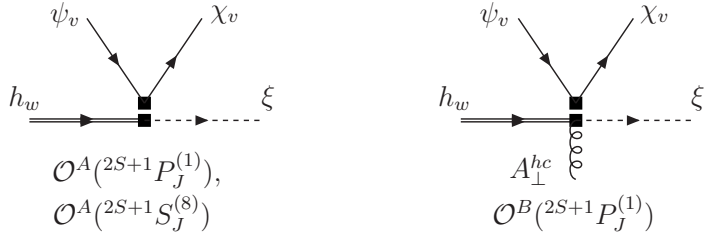}
  \vspace*{-22.8cm}
  \caption{Vertices corresponding to the $A$- and $B$-type operators.}
  \label{fig1}
  \end{center}
\end{figure}

Then, adopting the SCET notation defined in \cite{Beneke:2005vv},
we construct the colour-singlet operators
\bea
\nn
\mathcal{O}^A(^{2S+1}P_{J}^{(1)})&=&
\bigg[\tilde{C}_{f_+}^{(A0)}(\bar{\xi} W_c)\ns_+(1-\gamma_5) h_w
-\frac{1}{m_b} \int d\hat{s} \,\tilde{C}_{f_+}^{(B1)}(\hat{s})  \\
&&\hspace*{0.3cm}
(\bar{\xi} W_c)\ns_+[W^{\dagger}_c i \Ds_{\perp} W_c](s n_+)
(1+\gamma_5) h_w \bigg]
\,\mathcal{O}(^{2S+1}P_J^{(1)}),
\label{eq:5} \\[0.2cm]
\mathcal{O}^B(^{2S+1}\!P_{J}^{(1)})(s)&=& \frac{1}{m_b}
(\bar{\xi} W_c)\frac{\ns_+}{2}
[W^{\dagger}_c i \Ds_{\perp} W_c](s n_+)(1+\gamma_5) h_w
\,\mathcal{O}(^{2S+1}\!P_J^{(1)})
\label{btype}
\eea
associated with the vertex and spectator-scattering
amplitudes that have been considered in previous works~\cite{Song:2002mh,Song:2003yc,Pham:2005ih,Meng:2005fc,Meng:2005er,Meng:2006mi}.
The operators $\mathcal{O}^A(^{2S+1}\!P_J^{(1)})$ are written in such
a way that their tree-level matrix elements are proportional to the form
factor $f_+^{BK}(M_H^2)$ times the derivative of the
quarkonium wave function at the origin, since the expression
in square brackets in (\ref{eq:5}) is the SCET representation
of the full QCD form factor \cite{Beneke:2005gs}. The effective
vertices generated by the $A$- and $B$-type operators (in light-cone
gauge, where $W_c=1$)
are shown in Figure~\ref{fig1}.
New and central to the present discussion are the colour-octet operators
\bea
\nn
\mathcal{O}^A_\parallel(^{2S+1}S_{J}^{(8)})&=&
(\bar{\xi} W_c)\ns_+(1-\gamma_5) T^A h_w
\, \mathcal{O}(^{2S+1}S_J^{(8)}),
\\[0.2cm]
\mathcal{O}^A_{\perp}(^{2S+1}S_{J}^{(8)})&=&
(\bar{\xi} W_c)\gamma_{\perp \mu}(1-\gamma_5) T^A h_w
\, \mathcal{O}^{\mu}(^{2S+1}S_J^{(8)}).
\label{octetop}
\eea
According to the assumption that charmonium is a Coulomb bound
state, there exist two sets of low-energy scales, $m_c v, m_c v^2\gg
\Lambda$ related to the non-relativistic expansion and
$\sqrt{m_b\Lambda}$, $\Lambda$, related to the collinear expansion
and the strong-interaction scale. In the case of charmless decays
the matrix elements of colour-octet operators can be non-zero
only due to power-suppressed soft-gluon interactions, where soft
means momentum of order $\Lambda$, thus they
can be neglected at leading order in the $\Lambda/m_b$ expansion.
For charmonium, however, the decoupling of gluons with small momentum 
holds only when the momentum is much smaller than $m_c v^2$; 
gluons with momentum $m_c v^2$ contribute
to the octet operator matrix elements even at leading order in  $\Lambda/m_b$.
These contributions are subleading in $v$,
but so are the $P$-wave operators due to the extra derivative in
$\mathcal{O}(^{2S+1}P_J^{(1)})$,
hence the gluon-exchange contribution to the $S$-wave
octet operators is relevant at leading order in the velocity expansion 
to $P$-wave charmonium production.

Concentrating on the terms relevant to $P$-wave production at leading
(non-vanishing) order in the $v$ and $\Lambda/m_b$ expansion, the
effective weak-interaction Hamiltonian below the heavy quark mass
scale is therefore given by
\bea
\nn
\mathcal{H}_{\mbox{\scriptsize{eff}}} &=& \frac{G_F}{\sqrt{2}}
V_{cb}V^*_{cs} \,\sum_{S,J} \,\bigg\{
{\mathcal{C}}^A(^{2S+1}\!P_J^{(1)}) \,\mathcal{O}^A(^{2S+1}\!P_J^{(1)}) +
\int d \hat{s}\,\tilde{\mathcal{C}}^B(^{2S+1}\!P_J^{(1)})(\hat{s})\,
\mathcal{O}^B(^{2S+1}\!P_J^{(1)})(s) \\
&&\hspace*{0cm} +\,{\mathcal{C}}_\parallel^A(^{2S+1}S_J^{(8)})
\,\mathcal{O}_\parallel^A(^{2S+1}S_J^{(8)}) +
{\mathcal{C}}_{\perp}^A(^{2S+1}S_J^{(8)}) 
\,\mathcal{O}_{\perp}^A(^{2S+1}S_J^{(8)})  \bigg\},
\label{eq:6}
\eea
where we introduced the short-distance coefficients ${\mathcal{C}}$ and
$\hat{s}\equiv n_+ \cdot p \,s = M_B s$.

\subsection{\boldmath Tree-level matching of $A$-type operators}

The leading-order matching coefficients ${\mathcal{C}}^A$ are found
by comparing the $b\to c(p_c)\bar c(p_{\bar c}) s$ tree amplitude
(topology as in the left diagram of Figure~\ref{fig1})
computed with the effective Hamiltonian (\ref{eq:1}) to the
corresponding amplitude from (\ref{eq:6}). The result is:
\bea
\nn
\mathcal{C}_0^A(^1P_1^{(1)}) &=&  \mathcal{C}_0^A(^3P_0^{(1)}) =
\mathcal{C}_0^A(^3P_2^{(1)}) = 0,\\[0.15cm] \nn
\mathcal{C}_0^A(^3P_1^{(1)}) &=& -\frac{\sqrt{2}}{m_b\sqrt{z}}\left(C_2+\frac{C_1}{N_c}\right), \\ \nn
\mathcal{C}_\parallel^A(^1S_0^{(8)}) &=&
-\frac{1}{2\sqrt{z}}\left(2 C_1\right),
\qquad
\mathcal{C}_{\perp}^A(^1S_0^{(8)}) = - \left(2 C_1\right),\\
\mathcal{C}_\parallel^A(^3S_1^{(8)}) &=& \frac{1}{2}\left(2 C_1\right),
\qquad\quad\quad\,
\mathcal{C}_{\perp}^A(^3S_1^{(8)}) =\left(2 C_1\right).
\label{treematching}
\eea
Since the matrix elements of the colour-octet operators
and spectator-scattering are both suppressed by a factor of
$\alpha_s$, this reproduces the
well-known result from naive factorization that only
the $\chi_{c1}$ state is produced at leading order.

\subsection{Estimate of the branching fraction}

The leading-order decay amplitude is  now given by the expression
\begin{equation}
\mathcal{A}^{\mbox{{\scriptsize{LO}}}}_{B\rightarrow H(^{2S+1}P_J) K}
= \frac{G_F}{\sqrt{2}} \,V_{cb}V^*_{cs} \,
{\mathcal{C}}_0^A(^{2S+1}\!P_J^{(1)}) \,
\langle H(^{2S+1}P_J) \bar K |\mathcal{O}^A(^{2S+1}\!P_J^{(1)}) |
\bar B\rangle.
\end{equation}
The hadronic matrix element factorizes at leading order in the expansion 
in $\Lambda/m_b$ and $\alpha_s$ according to
\bea
\langle H(^{2S+1}P_J) \bar K |\mathcal{O}^A(^{2S+1}P_J^{(1)})|\bar B\rangle &=&
\langle \bar K |\bigg[\tilde{C}_{f_+}^{(A0)}(\bar{\xi} W_c)\ns_+(1-\gamma_5) 
h_w -\frac{1}{m_b} \int d\hat{s} \,\tilde{C}_{f_+}^{(B1)}(\hat{s})
\nonumber \\
&&\hspace*{-4.3cm}
(\bar{\xi} W_c)\ns_+[W^{\dagger}_c i \Ds_{\perp} W_c](s n_+)
(1+\gamma_5) h_w \bigg] | \bar B \rangle
\,\langle H(^{2S+1}P_J)|\mathcal{O}(^{2S+1}P_J^{(1)}) |0\rangle,
\quad
\label{mefact}
\eea
where the two factors reduce to the QCD $B\to K$ form factor,
\bea\nn
&&\langle \bar K |
\bigg[\tilde{C}_{f_+}^{(A0)}(\bar{\xi} W_c)\ns_+(1-\gamma_5) h_w
-\frac{1}{m_b} \int d\hat{s} \,\tilde{C}_{f_+}^{(B1)}(\hat{s}) \\
&&\qquad (\bar{\xi} W_c)\ns_+[W^{\dagger}_c i \Ds_{\perp} W_c](s n_+)
(1+\gamma_5) h_w \bigg] | \bar B \rangle =2 M_B \left(1-\frac{M_H^2}
{M_B^2}\right) f_+^{BK}(M_H^2),\quad
\label{eq:11b}
\eea
and the derivative of the charmonium wave function at the origin:
\bea\nn
\langle h_c |\mathcal{O}(^{1}P_1^{(1)}) |0 \rangle
&=&  n_-\cdot \varepsilon \,\langle \mathcal{O}(^{3}P_0^{(1)}) \rangle,
\\ \nn
\langle \chi_{c0} |\mathcal{O}(^{3}P_0^{(1)}) |0\rangle
&=&  \langle \mathcal{O}(^{3}P_0^{(1)}) \rangle,
\\ \nn
\langle \chi_{c1} |\mathcal{O}(^{3}P_1^{(1)}) |0\rangle
&=& n_-\cdot \varepsilon \, \langle \mathcal{O}(^{3}P_0^{(1)}) \rangle,
\\
\langle \chi_{c2} |\mathcal{O}(^{3}P_2^{(1)}) |0\rangle
&=&  n_-^{\mu}n_-^{\nu} \varepsilon_{\mu\nu} \,
 \langle \mathcal{O}(^{3}P_0^{(1)}) \rangle,
\label{eq:12}
\eea
with
\be
\label{metowf}
\langle\mathcal{O}(^{3}P_0^{(1)})\rangle =\sqrt{2 N_c} \sqrt{2M_{\chi_{c0}}}
\,(-i)\,\sqrt{\frac{3}{4\pi}} R^\prime_{21}(0).
\ee
Here spin symmetry of the leading non-relativistic interactions has been
used to write all four matrix elements in terms of
$\langle\mathcal{O}(^{3}P_0^{(1)})\rangle$, or equivalently
$R^\prime_{21}(0)$, where $R_{21}(r)$ denotes the radial
Schr\"odinger wave function of the $n=2$, $l=1$ $P$-wave
states, and the prime denotes a derivative.

Squaring the amplitude, integrating over the two-body phase space,
where we neglect the kaon mass, and summing over the charmonium
polarizations with the help of
\begin{equation}
\mbox{PolSum}[H(^{2S+1}P_J)] =
\left\{\,
\begin{array}{ll}
\,\,1 & H=\chi_{c0} \\[0.2cm]
\displaystyle \sum_{\lambda=-1}^1 \left|n_-\cdot \varepsilon(\lambda)\right|^2
= \frac{M_B^2}{M_H^2} & H=h_c, \chi_{c1}\\[0.2cm]
\displaystyle \sum_{\lambda=-2}^2 \left|n_-^{\mu}n_-^{\nu}
\varepsilon_{\mu\nu}(\lambda)\right|^2
= \frac{2 M_B^4}{3 M_{\chi_{c2}}^4} \qquad & H=\chi_{c2}
\end{array}
\right.
\label{polsum}
\end{equation}
we obtain the branching fraction
\begin{eqnarray}
\mbox{Br}\,(\bar B\rightarrow H(^{2S+1}P_J) \bar K)
&=& \tau_{B}\,\frac{G_F^2}{2}\,|V_{cb} V_{cs}^*|^2\,
\frac{M_B^2-M_H^2}{16\pi M_B^3}\,
\left|{\mathcal{C}}_0^A(^{2S+1}\!P_J^{(1)})\right|^2
\nonumber\\
&&\hspace*{-3cm} \times\,\mbox{PolSum}[H(^{2S+1}P_J)]\,
\left[2 M_B \left(1-\frac{M_H^2}
{M_B^2}\right) f_+^{BK}(M_H^2)\,\langle\mathcal{O}(^{3}P_0^{(1)})\rangle
\right]^2.
\label{brLO}
\end{eqnarray}

Due to (\ref{treematching}) this is different from zero at leading order
only for the $\chi_{c1}$ state. With parameters as given in 
Section~\ref{pheno}, varying the renormalization scale $\mu$
between 2 and 4.8~GeV, and $m_c$ between 1.45 and 1.75~GeV, we find
\begin{equation}
\mbox{Br}\,(\bar B\rightarrow \chi_{c1} \bar K)
= (0.13-1.06)\cdot 10^{-4}.
\end{equation}
The largest uncertainty arises from the scale-dependence of the
``colour-suppressed'' Wilson coefficient $C_2+C_1/N_c$.
The naive factorization prediction is at least a factor of four smaller
than the experimental result, see Table \ref{tab1}. 
This analysis shows that the dominant
contribution to the decay amplitude comes, for $P$-wave final state, from
(naively) non-factorizable dynamics, such as radiative corrections, 
spectator-scattering, and the colour-octet contributions.

\subsection{Overview of next-to-leading leading order terms}

The leading contribution to (\ref{eq:6}) comes from the colour-singlet
$P$-wave A-type operators $\mathcal{O}^A(^{2S+1}P_J^{(1)})$. Not
including the fields themselves in the
power-counting estimate, this contribution is of order $\alpha_s^0 v$,
where the factor $v$ arises from the derivatives in the
non-relativistic $P$-wave operators (\ref{eq:3}) and the
factor $\alpha_s^0$ is from the tree-level coefficient function.
Our intention is to show that radiative corrections
can be consistently computed, when $m_c v^2\gg \Lambda$, so we now
discuss the terms arising at order $\alpha_s v$:
\begin{itemize}
\item[1.] One-loop corrections
to the short-distance coefficients
$\mathcal{C}^A(^{2S+1}P_J^{(1)})$ of the $P$-wave colour-singlet
operators (``vertex corrections''):
\begin{equation}
\mathcal{A}^{\mbox{{\scriptsize{hard vertex}}}}_
{B\rightarrow H(^{2S+1}P_J) K} =
 \frac{G_F}{\sqrt{2}} \,V_{cb}V^*_{cs} \,
{\mathcal{C}}^A(^{2S+1}P_J^{(1)})|_{\mathrm{one-loop}} \,
\langle \mathcal{O}^A(^{2S+1}P_J^{(1)}) \rangle|_{\mathrm{tree}},
\label{hardvertex}
\end{equation}
where $\langle ...\rangle \equiv \langle H(^{2S+1}P_J^{(1)})\bar K|
... |\bar B \rangle$ and
$\langle \mathcal{O}^A(^{2S+1}P_J^{(1)}) \rangle|_{\mathrm{tree}}$
equals the factorized matrix element (\ref{mefact}).

Loop corrections to the matrix elements of these operators
need not be considered. In particular, corrections to the
factorization relation (\ref{mefact}) are suppressed by
powers of $\Lambda/m_b$, $v$ or $\alpha_s^2$, where the two
powers of the coupling arise from the vanishing colour-projection of
one-gluon exchange.

\item[2.] The tree-level matrix element of the B-type
$P$-wave colour-singlet operators \linebreak $\mathcal{O}^B(^{2S+1}P_J^{(1)})$
is of order $\alpha_s$, since one factor of $g_s$ is in the definition
of the operator and another is provided by the coupling of the
collinear gluon to the spectator-quark line (``spectator-scattering'').
The B-type form factor is of the same order as
$f_+^{BK}$~\cite{Beneke:2003pa}, hence this
contribution is also of order $\alpha_s v$. The hard
spectator-scattering amplitude is
\begin{equation}
\mathcal{A}^{\mbox{{\scriptsize{hard spectator}}}}_
{B\rightarrow H(^{2S+1}P_J) K} =
 \frac{G_F}{\sqrt{2}} \,V_{cb}V^*_{cs} \,
\int d \hat{s}\,\tilde{\mathcal{C}}^B(^{2S+1}P_J^{(1)})(\hat{s})|_
{\mathrm{tree}} \,
\langle \mathcal{O}^B(^{2S+1}P_J^{(1)})(s) \rangle|_{\mathrm{tree}}.
\label{hardspec}
\end{equation}

\item[3.] The colour-octet operators $\mathcal{O}^A(^{2S+1}S_J^{(8)})$
are of order $v^0$, but their
tree-level matrix elements vanish. The matrix element is non-zero
due to soft gluon exchange between the charm-quark lines and the
$b$-quark and light-quark lines, including the spectator
quark. At order $\alpha_s v$, one must
include the one-loop matrix element with exchange of a gluon of
momentum $m_c v^2$:
\begin{eqnarray}
\mathcal{A}^{\mbox{{\scriptsize{soft vertex}}}}_
{B\rightarrow H(^{2S+1}P_J) K} +
\mathcal{A}^{\mbox{{\scriptsize{soft spectator}}}}_
{B\rightarrow H(^{2S+1}P_J) K} &=&
\nonumber\\
&& \hspace*{-5cm}  \frac{G_F}{\sqrt{2}} \,V_{cb}V^*_{cs} \,
\sum_{a=\parallel,\perp} \sum_{J'=0,1}
{\mathcal{C}}_a^A(^{2S+1}S_{J'}^{(8)})|_{\mathrm{tree}} \,
\langle \mathcal{O}_a^A(^{2S+1}S_{J'}^{(8)}) \rangle|_{\mathrm{one-gluon}}.
\label{octetamplitude}
\end{eqnarray}
 Loop corrections to the short-distance coefficients
$\mathcal{C}^A(^{2S+1}S_{J'}^{(8)})$ need not be considered, since the
tree-level matrix element vanishes.

Colour-singlet $S$-wave operators contribute only at higher orders due 
to the vanishing colour projection in the one-gluon exchange contribution to 
their matrix elements.
\end{itemize}

\section{Short-distance contributions}

In this section we compute the short-distance contributions
related to items (\ref{hardvertex}), (\ref{hardspec}).

\subsection{One-loop short-distance coefficients}
\label{oneloopSD}

\begin{figure}[t]
  \vskip-1.1cm
  \begin{center}
  \hspace*{-2.2cm}
  \includegraphics[width=1.2\textwidth]{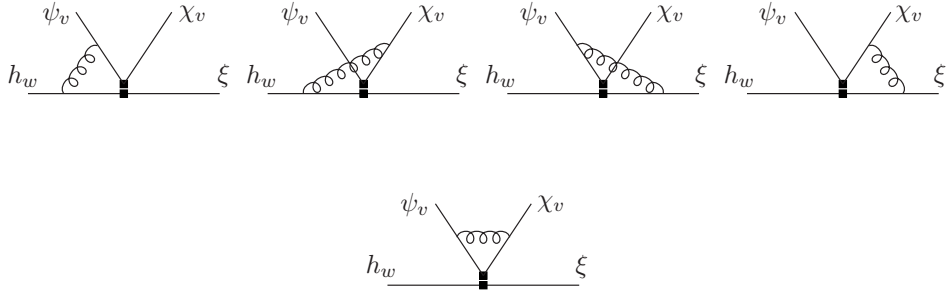}
  \vspace*{-22cm}
  \caption{One-loop corrections to the short-distance coefficients
  $\mathcal{C}^A(^{2S+1}P_J^{(1)})$ (vertex correction).}
  \label{fig2}
  \end{center}
\end{figure}

The one-loop correction to the short-distance coefficients of the
A-type colour-singlet $P$-wave operators (\ref{eq:5}) are obtained
from matching the QCD diagrams in Figure~\ref{fig2} to the
expression (\ref{hardvertex}). There is no contribution from the
diagram (not shown in the figure)
with gluon exchange between the $b$- and the $s$-quark line,
since the $B\to \chi_{c1} K$ amplitude at tree-level
involves only the form factor $f_+^{BK}$, which our operator
definition reproduces exactly, but not $f_0^{BK}$.

The loop correction is expected to be large, since it comes
with the large Wilson coefficient $C_1$, while the tree amplitudes
are either zero or colour-suppressed. We write
\bea
\nn
\mathcal{C}_{1}^A(\,^1\!P_1^{(1)}) &=&
\frac{1}{m_b \sqrt{z}}
\,\frac{\alpha_s C_F}{4\pi}\,\frac{C_1}{N_c}\,f[\,^1\!P_1],
\\
\nn
\mathcal{C}_{1}^A(\,^3\!P_0^{(1)}) &=&
-\frac{1}{\sqrt{3} m_b}
\,\frac{\alpha_s C_F}{4\pi}\,\frac{C_1}{N_c}\,f[\,^3\!P_0],
\\
\nn
\mathcal{C}_{1}^A(\,^3\!P_1^{(1)}) &=&
-\frac{\sqrt{2}}{m_b \sqrt{z}}
\,\frac{\alpha_s C_F}{4\pi}\,\frac{C_1}{N_c}
\left(f[\,^3\!P_1]-4\,\frac{N_c C_2+C_1}{C_1}\right),
\\
\mathcal{C}_{1}^A(\,^3\!P_2^{(1)}) &=&
-\frac{1}{m_b z}
\,\frac{\alpha_s C_F}{4\pi}\,\frac{C_1}{N_c}\,f[\,^3\!P_2],
\label{eq:13}
\eea
where the extra term in $\mathcal{C}_{1}^A(\,^3\!P_1^{(1)})$ arises from 
the $c\bar c$ vertex correction in the second line of Figure~\ref{fig2}.
The loop functions $f[\,^{2S+1}\!P_J]$ are extracted from
the $b(m_b w)\to c(p_c) \bar c(\bar p_c) s$ amplitude expanded
to first order
in the relative momentum $l$. The expansion is done in the integrand
to extract the hard momentum region, and the integration is performed
after expansion. The decomposition into
the four angular momentum states is done according to the operators
in (\ref{eq:3}). We use dimensional regularization with $d=4-2\epsilon$
for both ultraviolet and infrared singularities, and the NDR
scheme (naive anti-commuting $\gamma_5$) for the treatment of
$\gamma_5$ and the definition of the weak effective Hamiltonian
(\ref{eq:1}). Ultraviolet divergences are subtracted according to the
$\overline{\rm MS}$ prescription.
An ultraviolet divergence is present only in
$f[\,^3\!P_1]$, since for $^3P_1$ there exists a non-zero tree
amplitude, while infrared divergences appear for the other $P$-wave
operators but not for $^3\!P_1$. This can be understood from
the fact that the infrared divergences are related to the
$^1\!S_0$ and $^3\!S_1$ colour-octet matrix elements, and that soft gluons
change angular momentum by one unit, but do not change spin.

Defining the auxiliary function
\bea
f_{I0} &=&
-\left(\frac{1}{\epsilon}+\ln\frac{\mu^2}{m_b^2}\right)
\left(\frac{8}{1-z}+\frac{8\ln z}{(1-z)^2}\right) +
\frac{16\,\Big(\mbox{Li}_2\!\left[\frac{-1}{1-z}\right]-
\mbox{Li}_2\!\left[\frac{-z}{2(1-z)}\right]\Big)}{(1-z)^2}
\nonumber\\
&& -\,\frac{8 \ln^2 2}{(1-z)^2}+\frac{4\ln^2z}{(1-z)^2}
-\frac{16}{(1-z)^2}\,\ln\frac{2-z}{1-z}\,\ln \frac{z}{2}\,,
\label{eq:15}
\eea
and the ratio $z=4 m_c^2/m_b^2$ of heavy quark pole masses, we find
\bea \nn
f[\,^1\!P_1] &=&
z f_{I0}-\frac{4(2-3z+2z^2)}{(2-z)^2}
-\frac{4(8-8z+z^2+z^3)}{(2-z)^3}\left(\ln[1-z]-i\pi\right) \\ \nn
&&-\,\frac{4(6+16z-19z^2+5z^3)}{(1-z)(2-z)^2}\ln 2
-\frac{2z(8+13z-18z^2+5z^3)}{(1-z)^2(2-z)^2}\ln z, 
\\ \nn
f[\,^3\!P_0]&=&f_{I0}-\frac{4(2-7 z+2 z^2)}{z (2-z)^2}
-\frac{8(4-z-3 z^2+z^3)}{z (2-z)^3}\left(\ln[1-z]-i\pi\right) \\ \nn
&&-\,\frac{4(6+22z-27z^2+7z^3)}{z(1-z)(2-z)^2}\ln 2
+\frac{2(20-73z+58z^2-13z^3)}{(1-z)^2(2-z)^2}\ln z,
\\ \nn
f[\,^3\!P_1] &=& -6\ln\frac{\mu^2}{m_b^2} -18 +\frac{2 z}{2-z}
+\frac{4 (3-5z+2 z^2)}{(2-z)^2}\left(\ln[1-z]-i \pi\right) \\ \nn
&&-\,\frac{8z}{2-z}\ln 2 -\frac{4z^2}{(1-z)(2-z)}\ln z,
\\ \nn
f[\,^3\!P_2]&=&z^2 f_{I0}-\frac{4z(2-z-z^2)}{(2-z)^2}
-\frac{4z(8-8z+3z^2-z^3)}{(2-z)^3}\left(\ln[1-z]-i\pi\right)\\
&&-\,\frac{8z(3+8z-9z^2+2z^3)}{(1-z)(2-z)^2}\ln 2
-\frac{2z^2(16-5z-4z^2+z^3)}{(1-z)^2(2-z)^2}\ln z .
\label{eq:16}
\eea
The infrared $1/\epsilon$ pole, which violates factorization at the
heavy quark mass scale, is exhibited explicitly in $f_{I0}$. One should
note the difference with decays to light mesons or $S$-wave charmonia,
where the same hard vertex corrections are infrared finite. The
singularity arises here as a consequence of the expansion to first
order in the relative momentum $l$, and signals that -- unsurprisingly --
the colour-transparency argument does not hold at the heavy-quark
mass scale.

Results for the hard vertex correction, corresponding to the first four
diagrams in figure \ref{fig2}, have been obtained previously in
\cite{Song:2002mh,Song:2003yc,Pham:2005ih,Meng:2005fc,Meng:2005er,Meng:2006mi},
where the infrared divergences were noted for the first time. These 
papers use a gluon mass rather than space-time dimension as  
infrared regulator, while the finite part is given only in
parametric form. This makes difficult to compare these result with ours, 
except for the infrared divergent part, where we agree, and for the 
$\chi_{c1}$ case, where the finite contribution is given explicitly 
\cite{Meng:2005er} and 
there is no infrared regulator dependence, 
and where we also agree. 
The fifth diagram in Figure~\ref{fig2} was not 
taken into account in previous papers.

\subsection{Short-distance spectator scattering}
\label{sec2.3}

\begin{figure}[t]
  \vskip-0.8cm
  \begin{center}
  \hspace*{-5.3cm}
  \includegraphics[width=1.4\textwidth]{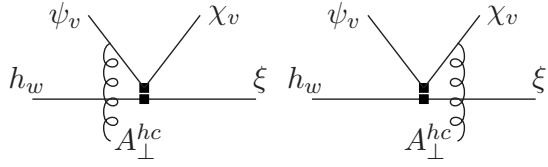}
  \vspace*{-28cm}
  \caption{Hard spectator-scattering -- tree diagrams for the
  coefficient functions $\mathcal{C}_{1}^B(^{2S+1}\!P_J)(y)$.}
  \label{fig3}
  \end{center}
\end{figure}

The tree contribution to the short-distance coefficients of the
$B$-type colour-singlet $P$-wave operators (\ref{btype}) is obtained
from matching the QCD amplitude for the process $b(m_b w)\to
c(p_c) \bar c(p_{\bar c}) s(y p) g(\bar y p)$ shown
in Figure~\ref{fig3} to the expression (\ref{hardspec}).
In this way, we calculate directly the momentum-space
coefficient function
\be
{\mathcal{C}}^B(\,^{2S+1}P_J^{(1)})(y) = \int d\hat{s}\,
e^{i \bar{y} \hat{s}}\,\tilde{\mathcal{C}}^B(\,^{2S+1}\!P_J^{(1)})(\hat{s}),
\label{eq:22}
\ee
where $\bar y\equiv 1-y$ is the fraction of longitudinal momentum
$n_+ p = m_b (1-z)$ carried by the hard-collinear gluon. We find
\bea
\mathcal{C}_{1}^B(^1P_1^{(1)})(y) &=&
-\frac{2C_1}{N_c}\frac{2}{m_b \sqrt{z}},
\nonumber \\
\nn
\mathcal{C}_{1}^B(^3P_0^{(1)})(y) &=&
\frac{2C_1}{N_c}\frac{2}{\sqrt{3}m_b z}
\left(1-\frac{2z}{1-z}\frac{1}{\bar{y}}\right),
\\
\nn
\mathcal{C}_{1}^B(^3P_1^{(1)})(y) &=&
-\frac{2C_1}{N_c}\frac{2\sqrt{2}}{m_b \sqrt{z}}
\left(1+\frac{z}{1-z}\frac{1}{\bar{y}}\right),
\\
\mathcal{C}_{1}^B(^3P_2^{(1)})(y) &=&  \frac{2C_1}{N_c}
\frac{2}{m_b}\left(1+\frac{z}{1-z}\frac{1}{\bar{y}}\right).
\label{eq:23}
\eea
For later purposes it will be useful to express these results as
\begin{equation}
{\mathcal{C}}^B(\,^{2S+1}\!P_J^{(1)})(y) =
\frac{2C_1}{N_c} \left(C^B[\,^{2S+1}\!P_J\,]+
\frac{B[\,^{2S+1}\!P_J\,]}{\bar y}\right)
\label{cb1}
\end{equation}
with $y$-independent coefficients $C^B[\,^{2S+1}P_J\,]$,
$B[\,^{2S+1}P_J\,]$, which follow by comparison with
(\ref{eq:23}).

\begin{figure}[t]
  \vskip-0.8cm
  \begin{center}
  \hspace*{-2.3cm}
  \includegraphics[width=1.3\textwidth]{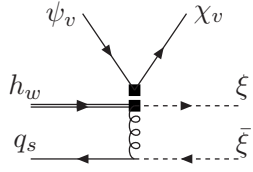}
  \vspace*{-26cm}
  \caption{Hard spectator-scattering -- matrix element.}
  \label{fig4}
  \end{center}
\end{figure}
The hard spectator-scattering amplitude (\ref{hardspec})
requires the evaluation of the convolution
\begin{equation}
\int d \hat{s}\,\tilde{\mathcal{C}}^B(^{2S+1}P_J^{(1)})(\hat{s})|_
{\mathrm{tree}} \,
\langle \mathcal{O}^B(^{2S+1}P_J^{(1)})(s) \rangle|_{\mathrm{tree}},
\label{hardspecconv}
\end{equation}
which is done following \cite{Beneke:2005vv}. The corresponding
diagram is shown in Figure~\ref{fig4}. First, as in the case
of the A-type operators, the tree matrix element factorizes
according to
\bea
\nn
\langle H(^{2S+1}P_J) \bar K |\mathcal{O}^B(^{2S+1}P_J^{(1)})(s)|
\bar B\rangle &=&
\langle \bar K |\frac{1}{m_b}(\bar{\xi} W_c)\frac{\ns_+}{2}
[W^{\dagger}_c i \Ds_{\perp} W_c](s n_+)(1+\gamma_5) h_w|\bar B\rangle \\
&&\hspace*{-4cm} \times\,
\langle H(^{2S+1}P_J)|\mathcal{O}(^{2S+1}P_J^{(1)}) |0\rangle.
\label{eq:17}
\eea
The heavy-to-light matrix element is the same as appears in non-leptonic
decay to two light mesons resulting in (\cite{Beneke:2005vv}, Eqs.~(6),~(8))
\bea
\nn
&&\langle \bar K |\frac{1}{m_b}(\bar{\xi} W_c)\frac{\ns_+}{2}
[W^{\dagger}_c i \Ds_{\perp} W_c](s n_+)(1+\gamma_5) h_w|\bar B\rangle
= \\
&&\hspace*{2.0cm} -\frac{2 E_K M_B}{4 m_b}
\int_0^1 d\tau \int_0^{\infty} d\omega \int_0^1 d y \,
e^{i \bar{\tau} \hat s } J_{\parallel}(\bar{\tau};y,\omega)
f_B \, \phi_{B+}(\omega) f_K \, \phi_{\bar K}(y).
\label{eq:18}
\eea
Here  $\phi_{B+}(\omega)$ and $\phi_{\bar K}(y)$ are the leading-twist
light-cone distribution amplitudes of the $B$ and the $\bar K$ meson,
respectively, while the hard-collinear
``jet function" $J_{\parallel}(\bar{\tau};y,\omega)$ is given at
leading order by
\be
J^{(0)}_{\parallel}(\bar{\tau};y,\omega)=-\frac{4 \pi \alpha_s C_F}{N_c}
\frac{1}{2 E_K\omega \bar{y}}
\,\delta(\bar{\tau}-\bar{y}).
\label{eq:19}
\ee
Introducing
\be
\label{eq:24}
\int_0^{\infty} d\omega \,\frac{\phi_{B+}(\omega)}{\omega}\equiv
\frac{1}{\lambda_B},
\ee
the tree-level matrix elements reads
\bea
\nn
&&\langle \bar K |\frac{1}{m_b}(\bar{\xi} W_c)\frac{\ns_+}{2}
[W^{\dagger}_c i \Ds_{\perp} W_c](s n_+)(1+\gamma_5) h_w|
\bar B\rangle_{\rm tree}
= \\
&&\hspace*{2.0cm} \frac{\pi \alpha_s C_F}{N_c}\,
\frac{f_B M_B}{m_b \lambda_B}
\int_0^1 d y \,e^{i \bar{y} \hat s }\,\frac{1}{\bar y}
\,f_K \phi_{\bar K}(y).
\label{btypeff}
\eea
Inserting the Fourier representation of the coefficient function
into (\ref{hardspecconv}), we find the amplitude
\begin{eqnarray}
\mathcal{A}^{\mbox{{\scriptsize{hard spectator}}}}_
{B\rightarrow H(^{2S+1}P_J) K} &=&
\frac{G_F}{\sqrt{2}} \,V_{cb}V^*_{cs} \,
\frac{\pi \alpha_s C_F}{N_c}\,
\langle H(^{2S+1}P_J)|\mathcal{O}(^{2S+1}P_J^{(1)}) |0\rangle
\nonumber\\
&& \times\,\frac{f_K f_B M_B}{m_b \lambda_B}
\int_0^1 d y \,{\mathcal{C}}^B(\,^{2S+1}P_J^{(1)})(y)\,
\frac{\phi_{\bar K}(y)}{\bar y}.
\label{hardspecamp1}
\end{eqnarray}
This result has been obtained previously 
\cite{Song:2002mh,Song:2003yc,Pham:2005ih,Meng:2005fc,Meng:2005er}
by direct evaluation of the spectator-scattering amplitude.

The problematic aspect of this expression is the endpoint
divergence of the convolution integral when $y\to 1$. The kaon
light-cone distribution amplitude behaves as $\phi_{\bar K}(y)
\propto \bar y$ for small $\bar y$. But contrary to the situation
for decays to two light mesons or an $S$-wave charmonium and
a kaon, the coefficient function ${\mathcal{C}}^B(\,^{2S+1}P_J^{(1)})(y)$
contains a piece proportional to $1/\bar y$ resulting in a
logarithmically divergent integral. We regularize this integral
by introducing a cutoff that replaces the upper limit by
$1-\mu$ with $\mu\ll 1$. Using
\begin{equation}
\int_0^{1-\mu} d y \,\frac{\phi_{\bar K}(y)}{\bar y^2}
= \phi_{\bar K}^\prime(1) \ln\mu+
\int_0^1 dy\,\frac{\phi_{\bar K}(y)+\bar y \,
\phi_{\bar K}^\prime(1)}{\bar y^2}
+O(\mu)
\label{subtractend}
\end{equation}
the convolution integral in (\ref{hardspecamp1}) takes the
final form
\begin{eqnarray}
&& \int_0^{1-\mu} d y \,{\mathcal{C}}^B(\,^{2S+1}P_J^{(1)})(y)\,
\frac{\phi_{\bar K}(y)}{\bar y} =
\frac{2 C_1}{N_c} \,\Bigg\{
B[\,^{2S+1}\!P_J\,] \,\phi_{\bar K}^\prime(1) \ln\mu
\nonumber\\
&& \hspace*{2cm}+ \, B[\,^{2S+1}\!P_J\,]
\int_0^1 dy\,\frac{\phi_{\bar K}(y)+\bar y \,
\phi_{\bar K}^\prime(1)}{\bar y^2}
+ C^B[\,^{2S+1}\!P_J\,]
\int_0^{1} d y \,\frac{\phi_{\bar K}(y)}{\bar y}
\Bigg\}.\quad
\label{finalconv}
\end{eqnarray}
The regulator-dependent $\ln\mu$ term appears to violate factorization
even at leading power in the heavy quark expansion.
We shall show below, however, that this dependence is canceled
by a corresponding ultraviolet divergence in the colour-octet
matrix elements.

\section{Colour-octet matrix elements}
\label{sec3}

The important new element in our treatment of $B$ decays to $P$-wave
charmonia are the colour-octet contributions (\ref{eq:6}),
(\ref{octetamplitude}) to the decay amplitude. We shall now show
how to compute these
matrix elements when the charmonium is a Coulomb bound state and
demonstrate that the infrared singularities in the vertex correction
and hard spectator-scattering can be absorbed into a renormalization
of the colour-octet matrix elements. In the Coulomb limit the
octet matrix elements can be computed in perturbation theory, and
we provide results at order $\alpha_s$, corresponding to the accuracy
of the short-distance terms. In general, and more realistically,
the colour-octet matrix elements may be introduced as new non-perturbative
parameters, but their scale dependence still cancels the 
factorization scale dependence
of the hard-scattering terms. However, factorization in the sense
of separating the $B\to K$ transition from the vacuum to
charmonium matrix element holds only in the Coulomb limit, since
otherwise the octet matrix elements of the SCET/NRQCD four-quark operators
contain strongly interacting gluon exchanges between the charmonium
and the $BK$ system.

\subsection{Reduction formula for quarkonium
matrix elements}

We briefly review the formalism for the calculation of quarkonium
matrix elements
\begin{equation}
\langle H X|{\cal O}|Y\rangle,
\label{Hamp}
\end{equation}
where ${\cal O}$ is some operator, $H$ the quarkonium  with momentum
$q=M_H v$ in some polarization state, and $X$,$Y$ denote arbitrary other
particles.

\begin{figure}[t]
  \vskip-0.8cm
  \begin{center}
  \hspace*{-2.3cm}
  \includegraphics[width=1.3\textwidth]{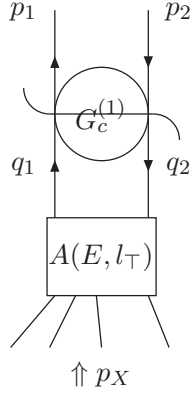}
  \vspace*{-23.1cm}
\caption{Amplitude exhibiting a quarkonium pole after summing Coulomb
gluon exchanges in $G_c^{(1)}$.}
\label{fig5}
\end{center}
\end{figure}

The quarkonium matrix element is identified in the standard
way as part of the residue factor of the pole of a suitable Green function.
We pick a Green function with two external charm quark fields. The
quarkonium bound state poles appear after summing ladder diagrams
with colour-singlet Coulomb (potential) gluon exchange as the
bound state poles of the Coulomb Green function $G_c^{(1)}$ as
indicated in Figure~\ref{fig5}. For external momenta
$p_1, p_2$ with $(p_1+p_2)^2$ near $4 m_c^2$ the
bound state poles appear in
the integration region where $q_1=m_c v + \ell_1$, $q_2 = m_c v+ \ell_2$
with $E\equiv v\cdot (q_1+q_2)-2 m_c = v \cdot (\ell_1+\ell_2)$
and $\ell_{1\top}$, $\ell_{2\top}=-\ell_{1\top}\equiv -\ell_\top$
small, of order $m_c\,\alpha_s^2$
and $m_c\,\alpha_s$, respectively. A specific charmonium state is
extracted by performing the following steps: (1) Insert the
spectral representation of the Green function and use that near
a bound state pole
\be
\tilde{G}_c^{(1)}(\bfm{k},\bfm{\ell}; E)
\stackrel{E\rightarrow E_n}{\rightarrow}
\sum_{l,m} \frac{\Psi_{nlm}(\bfm{k})\Psi^{*}_{nlm}(\bfm{\ell})}
{E_n-E-i\epsilon},
\ee
where $\Psi_{nlm}(\bfm{\ell})$ is the momentum space wave function
of the Schr\"odinger operator
in the $nlm$ basis. The three-vectors are introduced by
writing four-vectors $a_\top^\mu$ orthogonal to $v$ as
$a_\top^\mu = L^\mu_{\phantom{\mu}\nu}(v) \,\hat a^\nu$,
where $\hat a^\mu=(0,\bf{a})$
and $L^\mu_{\phantom{\mu}\nu}(v)$ is the Lorentz boost
from the quarkonium rest
frame to the frame where its momentum is $M_H v$.
(2) To separate the degenerate states with different angular momentum
and spin, first introduce the spherical decomposition
\begin{equation}
\Psi_{nlm}(\bfm{\ell}) = \tilde R_{nl}(\ell) \,Y_{lm}(\Omega)
\end{equation}
defining $\ell=|\bfm{\ell}|$.
Then perform a Fierz transformation in the Dirac indices of the
two intermediate charm propagators in Figure~\ref{fig5} to obtain
the projection on the spin zero and spin 1 components. For $l=1$,
the $J=0,1,2$ states follow from the standard Clebsch-Gordon relations.

The final result can be expressed in terms of the on-shell $c\bar c$
matrix element corresponding to (\ref{Hamp}), which we
write in the from
\begin{equation}
\langle c(q_1) \bar c(q_2) X|{\cal O}|Y\rangle =
\bar u_c(q_1) A(E,\ell_\top)\, v_c(q_2).
\end{equation}
Defining the matrices
\begin{equation}
\Lambda[H] = \left\{\hspace*{0.2cm}
\begin{array}{ll}
\displaystyle
- \frac{\epsilon^*(\lambda)\cdot \ell_\top}{\ell} \,\gamma_5
& H =h_c \\[0.3cm]
\displaystyle
\frac{1}{\sqrt{3}}\,\frac{\not\!\ell_\top}{\ell}
& H=\chi_{c0} \\[0.3cm]
\displaystyle \frac{1}{2\sqrt{2}}\,\Big[\frac{\not\!\ell_\top}{\ell},
\not \!\epsilon^*(\lambda)\Big]\,\gamma_5\qquad
& H=\chi_{c1}\\[0.3cm]
\displaystyle
-\epsilon^*_{\alpha\beta}(\lambda) \,\frac{\ell_\top^\alpha}{\ell}
\,\gamma_\top^\beta
& H=\chi_{c2}
\end{array}
\right.
\end{equation}
in Dirac-index space (and diagonal in colour space), the desired
quarkonium matrix element is
\begin{equation}
\langle H X|{\cal O}|Y\rangle =
\sqrt{2 M_H}\,\frac{1}{\sqrt{2 N_c}}
\int \frac{d^3\bfm{\ell}}{(2\pi)^3}\,\sqrt{\frac{3}{4\pi}}\,
\tilde R_{21}(\ell)\,\,
\mbox{tr}\left(\Lambda[H]\,\frac{1+\not\!v}{2}\, A(E,\ell_\top)
\,\frac{1-\not\!v}{2}\right).
\label{quarkoniumLSZ}
\end{equation}
This is valid in a leading-order treatment of the non-relativistic bound
state dynamics. Beyond this approximation, corrections to the wave-function
and trace expression are required. Eq.~(\ref{quarkoniumLSZ}) can be used
to calculate the first non-vanishing contribution to a quarkonium
matrix element, and this will be sufficient for the colour-octet
terms considered below.

The momentum-space radial wave function follows from the Fourier transform
of the position-space expression $R_{nl}(r) Y_{lm}(\Omega)$ and is given
by
\begin{equation}
\tilde R_{nl}(k) = (-i)^l\int_0^\infty dr\,4\pi r^2\,j_l(k r)
R_{nl}(r).
\end{equation}
For the case $n=2$, $l=1$, using the spherical Bessel function
\begin{equation}
j_1(kr) = \frac{\sin kr - kr \cos kr}{(kr)^2},
\end{equation}
the integral evaluates to
\begin{equation}
\tilde R_{21}(k) = (-i) \,\frac{1024\pi \gamma_B k}{(4 k^2+\gamma_B^2)^3}
\,R^\prime_{21}(0),
\label{momwf}
\end{equation}
where $R^\prime_{21}(0)$ denotes the derivative of the position-space
wave function at the origin, and
\begin{equation}
\gamma_B = \frac{m_c\alpha_s C_F}{2}
\label{bohrradius}
\end{equation}
is the inverse Bohr radius of the charmonium.

As an example of using (\ref{quarkoniumLSZ}) we calculate
$\langle \chi_{c1}|\mathcal{O}(^3P_1^{(1)})|0\rangle$ with
$\mathcal{O}(^3P_1^{(1)})$ given in (\ref{eq:3}). The tree amplitude is
\begin{equation}
A(E,\ell_\top) = \frac{n_{-\mu}}{2\sqrt{2}}
\left[\not\!\ell_{\top},\gamma_{\top}^{\mu}
\right]\gamma_5
\end{equation}
resulting in
\begin{eqnarray}
\mbox{tr}\left(\Lambda[\chi_{c1}]\,\frac{1+\not\!v}{2}\, A(E,\ell_\top)
\,\frac{1-\not\!v}{2}\right) =
\frac{2 N_c}{3} \,\ell \,n_-\cdot\epsilon(\lambda).
\end{eqnarray}
It follows that
\begin{eqnarray}
\langle \chi_{c1}|\mathcal{O}(^3P_1^{(1)})|0\rangle &=&
\sqrt{2 N_c} \sqrt{2 M_{\chi_{c1}}}\,n_-\cdot\epsilon(\lambda)\,\frac{1}{3}
\int \frac{d^3\bfm{\ell}}{(2\pi)^3}\,\sqrt{\frac{3}{4\pi}}\,
\tilde R_{21}(\ell)\,\ell
\nonumber\\
&=&
n_-\cdot\epsilon(\lambda)\,\sqrt{2 N_c} \sqrt{2 M_{\chi_{c1}}}\,
\sqrt{\frac{3}{4\pi}}\,(-i)\,R^\prime_{21}(0),
\end{eqnarray}
which is consistent with (\ref{eq:12}), (\ref{metowf}).

\subsection{Soft vertex correction}

We proceed to the calculation of the colour-octet matrix elements
$\langle \chi_{cJ} \bar K| \mathcal{O}^A_a
(^{3}S_1^{(8)})|\bar B \rangle$ and $\langle h_c \bar K| \mathcal{O}^A_a
(^{1}S_0^{(8)})|\bar B \rangle$ ($a=\,\parallel,\perp$).
Note that at order $\alpha_s$ the interactions are spin-symmetric,
so there is no contribution of the $^{3}S_1^{(8)}$ ($^{1}S_0^{(8)}$)
operator to the $h_c$ ($\chi_{cJ}$) final state. Each matrix
element receives contributions from vertex diagrams (first four diagrams
in Figure~\ref{fig2}) and spectator-scattering diagrams (Figure~\ref{fig3}
with gluon attached to the undisplayed spectator quark line), except
that now the gluon virtuality is small, of order $(m_c v^2)^2$ for 
the vertex diagrams.

We begin by writing down the soft gluon coupling to the $c\bar c$ pair.
The leading interactions of dynamical gluons with momentum of
order $m_c v^2$ to the heavy charm quarks are provided by the (P)NRQCD
effective Lagrangian. They read $g_s \psi^\dagger(x) (A^0(t,\bfm{0}) -
\bfm{x}\cdot \bfm{E}(t,\bfm{0}))\psi(x)$ together with a similar term
for the antiquark field. The contribution from the $A^0$ coupling
cancels in the sum of the attachments to the $c$ and the $\bar c$
line (or can be gauged away), leaving the chromoelectric dipole
interaction. The dipole interaction provides the additional factor
of velocity $v$ which renders the octet $S$-wave operator
matrix element of the same order in $v$ as the singlet $P$-wave
operator.

\begin{figure}[t]
  \vskip-0.8cm
  \begin{center}
  \hspace*{-2.3cm}
  \includegraphics[width=1.3\textwidth]{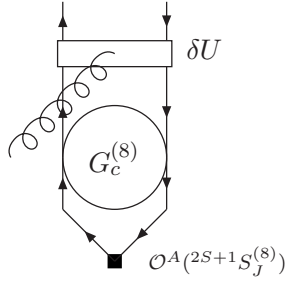}
  \vspace*{-25cm}
\caption{Soft gluon attachment to the charm quark lines.}
\label{fig6}
\end{center}
\end{figure}

The part of the amplitude involving the charm quark lines can now be
expressed in the form
\begin{equation}
 A_{c\bar c}(E,\ell_\top) = \int\frac{d^3 \ell^\prime_\top}{(2\pi)^3}\,
\Gamma_a[^{2S+1}S_J^{(8)}]\,
i G_c^{(8)}(\ell^\prime _{\top},k_\top+\ell_{\top}; E + v\cdot k)
\,\delta U (k+\ell,\ell),
\label{ccamp}
\end{equation}
see Figure~\ref{fig6}. The various items in this equation are as follows:
$k$ denotes the outgoing soft gluon momentum; $\Gamma_a[^{2S+1}S_J^{(8)}]$
comes from the $c\bar c$ part of the colour-octet operator as
given by the contents of square brackets in (\ref{eq:3});
$\delta U (k+\ell,\ell)$ is the momentum-space soft gluon interaction
vertex
\be
\delta U (k+\ell,\ell) = -i g_s T^B
\stackrel{\longleftarrow}{\frac{\partial}{\partial \ell_{\top}^{\lambda}}}
\left(v^{\rho} k_{\top}^{\lambda} - g_\top^{\rho\lambda} v\cdot k
\right),
\label{usvertex}
\ee
where $\rho$ and ``B'' are, respectively,
the Lorentz and colour index of the soft gluon;
$G_c^{(8)}(\ell^\prime _{\top},k_\top+\ell_{\top}; E + v\cdot k)$
is the Coulomb Green function that sums an infinite number of gluon
exchanges ``between'' the operator and the soft gluon vertex. Here
we need the colour-octet Green function, since the $c\bar c$ pair is
produced at the operator vertex in a colour-octet state. The calculation
of the vertex diagrams with the full Coulomb Green function is
quite involved (see \cite{Beneke:1999gq} for the calculation of
colour-octet inclusive quarkonium production matrix elements), but turns
out to be unnecessary to good approximation. Similar calculations
involving colour-octet Coulomb Green
functions \cite{Beneke:2007pj,Beneke:2008cr} find that the
numerically largest term arises from the no-gluon exchange
term, since every colour-octet exchange is suppressed by the
small colour factor $-1/(2 N_c)$. We therefore simplify (\ref{ccamp})
by approximating
\be
i G_c^{(8)}(\ell_{\top},\ell_{\top}';E) =
\frac{-i}{E+\frac{\ell_{\top}^2}{m_c}+i\epsilon}
\,(2\pi)^3 \delta^{(3)}(\ell_{\top}-\ell'_{\top}),
\label{octetG}
\ee
obtaining
\begin{equation}
 A_{c\bar c}(E,\ell_\top) = i^2 g_s\,
\Gamma_a[^{2S+1}S_J^{(8)}]\,
\left(-\frac{2}{m_c}\right)\,
\frac{\left(v^{\rho} k_{\top}^{\lambda} - g_\top^{\rho\lambda} v\cdot k\right)
\ell_{\top\lambda}}
{\displaystyle
\phantom{E}\bigg(E+v\cdot k+\frac{\ell_{\top}^2}{m_c}+i\epsilon\bigg)^{\!2
\phantom{\vec{E}}}} \, T^B.
\label{ccamp2}
\end{equation}

Attaching the gluon to the bottom quark and the strange quark line and
making use of (\ref{quarkoniumLSZ}), we arrive at
\begin{eqnarray}
\langle H \bar K| \mathcal{O}^A_\parallel(^{2S+1}S_J^{(8)})|
\bar B \rangle|_{\mbox{{\scriptsize vertex}}}
&=&
\sqrt{2 M_H}\,\frac{1}{\sqrt{2 N_c}}
\int \frac{d^3\bfm{\ell}}{(2\pi)^3}\,\sqrt{\frac{3}{4\pi}}\,
\tilde R_{21}(\ell)\,\tilde\mu^{2\epsilon} \int \frac{d^dk}{(2\pi)^d}
\frac{(-i)}{k^2+i\epsilon}
\nonumber\\
&& \hspace*{-2cm}
\times \,
\mbox{tr}\left(\Lambda[H]\,\frac{1+\not\!v}{2}\, A_{c\bar c}(E_H,\ell_\top)
\,\frac{1-\not\!v}{2}\right)
\nonumber\\
&& \hspace*{-2cm}
\times \,
i^2 g_s \left(\frac{p_{B\rho}}{p_B\cdot k} -
\frac{p_{\rho}}{p\cdot k}\right)
\langle \bar K |(\bar{\xi} W_c)
\!\not\!n_+ (1-\gamma_5) T^A T^B h_w|\bar B\rangle,
\label{octetvertex1}
\end{eqnarray}
where we used that $k$ is soft to simplify the bottom and strange quark
propagators, and denote the binding energy by
\begin{equation}
E_H = M_H-2m_c+i\epsilon.
\label{bindingenergy}
\end{equation}
The colour index ``A'' is contracted with an index hidden in the definition
of $\Gamma_a[^{2S+1}S_J^{(8)}]$, while $\rho$ and ``B'' are contracted
with the corresponding indices in (\ref{ccamp2}). The order of the
colour matrices $T^A T^B$, which is different for the attachment to the
bottom and strange line plays no role, since the colour part of
the trace evaluates to $\frac{1}{2}\delta^{AB}$. The corresponding
matrix elements of the operators $\mathcal{O}^A_\perp(^{2S+1}S_J^{(8)})$
have $\gamma_{\perp\mu}$ instead of $\!\not\!n_+$ in the
$\langle \bar K | ... |\bar B\rangle$ matrix element. The
absence of a transverse vector implies that the matrix element vanishes,
so at this order
\begin{equation}
\langle H \bar K| \mathcal{O}^A_\perp(^{2S+1}S_J^{(8)})|
\bar B \rangle|_{\mbox{{\scriptsize vertex}}} =0.
\end{equation}

The integrations appearing in (\ref{octetvertex1}) can be performed
exactly. The loop integral is of the general form
\begin{equation}
\tilde\mu^{2\epsilon} \int \frac{d^dk}{(2\pi)^d}
\,\frac{k_\mu}{k^2 P\cdot k \,(v\cdot k +\omega)^2} =
\frac{i}{(4\pi)^2}\left(I_P \,v_\mu + J_P \,P_\mu\right).
\end{equation}
Only $I_P$ enters the final result, and can be calculated directly 
by introducing Feynman parameters. In contrast to the hard
vertex correction, this loop integral is infrared finite. The sensitivity
to very soft gluon momenta is cut off at the scale of the binding
energy $E_B$ since $\omega =E_B-\ell_\top^2/m_c$.
On the other hand, there is a logarithmic ultraviolet
divergence, which we regulate dimensionally to be consistent with
the calculation of the short-distance correction. The angular
$\bfm{\ell}$-integral is also easily done 
using\footnote{Note $\ell_\top^2=-\bfm{\ell}^2=-\ell^2$.}
\begin{equation}
\int d\Omega \,\ell_\top^\alpha \ell_\top^\beta = -\frac{4\pi}{3}\,
\ell^2\,g_\top^{\alpha\beta}.
\end{equation}
The remaining $\ell$-integral is of the type
\begin{equation}
\int_0^\infty d\ell \,\ell^3\,\tilde R_{21}(\ell)
\left[A+B\ln\left(-\frac{2 (E_H-\ell^2/m_c)}{\mu}\right)\right]
\label{lint}
\end{equation}
with $\ell$-independent constants $A$, $B$, and is evaluated using
the explicit form (\ref{momwf}) of the momentum-space
radial wave function.
Including the colour and spin traces we find
\begin{eqnarray}
&& \langle h_c \bar K| \mathcal{O}^A_\parallel(^{1}S_0^{(8)})|
\bar B \rangle|_{\mbox{{\scriptsize vertex}}}
= -\frac{\alpha_s}{4\pi}\,\frac{C_F}{N_c}\,
{\cal D}_{sv}(\,^1\!P_1^{(1)})\,F_{sv}\,
\langle h_c \bar K |\mathcal{O}^A(^{1}P_1^{(1)})
|\bar B\rangle|_{\mbox{{\scriptsize tree}}},
\nonumber
\\
&&\langle \chi_{cJ} \bar K| \mathcal{O}^A_\parallel(^{3}S_1^{(8)})|
\bar B \rangle|_{\mbox{{\scriptsize vertex}}}
= -\frac{\alpha_s}{4\pi}\,\frac{C_F}{N_c}\,
{\cal D}_{sv}(\,^3\!P_J^{(1)})\,F_{sv}\,
\langle \chi_{cJ} \bar K |\mathcal{O}^A(^{3}P_J^{(1)})
|\bar B\rangle|_{\mbox{{\scriptsize tree}}},
\quad
\label{octetvertex2}
\end{eqnarray}
where we have re-expressed the product of the derivative of the wave
function at the origin and the $B\to K$ form factor in terms of
the factorized matrix element (\ref{mefact})
to facilitate the comparison
with the hard vertex amplitude (\ref{hardvertex}). Eq.~(\ref{octetvertex2})
contains the spin-dependent coefficients
${\cal D}_{sv}(\,^1\!P_1^{(1)}) =1$ and
${\cal D}_{sv}(\,^3\!P_J^{(1)})=\{1/(\sqrt{3} z),0,1\}$ (for
$J=0,1,2$), and the loop function
\begin{eqnarray}
F_{sv} &=& \frac{8 z}{m_b (1-z)^2}
\Bigg(\left(1-z+\ln z\right)
\Bigg[\frac{1}{\epsilon}+\ln\frac{\mu^2}{m_b^2}+\ln z
-2 \Delta F\Bigg]
\nonumber\\
&& +2\,\mbox{Li}_2(z)-\frac{1}{2}\ln^2 z+2\ln z \ln(1-z) -2 \ln z-
\frac{\pi^2}{3}\Bigg),
\nonumber\\
\Delta F &=& \ln \frac{(2\sqrt{-m_c E_H}+\gamma_B)^2}{m_b^2} +
\frac{4}{3}\,\frac{\gamma_B(5\sqrt{-m_c E_H}+2\gamma_B)}
{(2\sqrt{-m_c E_H}+\gamma_B)^2}.
\label{octetloopfn}
\end{eqnarray}
This includes the ultraviolet divergence explicitly, as well as 
dependence on the
bound state parameters $\gamma_B$ (\ref{bohrradius}) and $E_H$
(\ref{bindingenergy}) in $\Delta F$. The colour-octet matrix elements
are complex, since they contain soft-rescattering phases. However,
as can be seen from the expression for $\Delta F$ or from
(\ref{lint}), in the present one-loop approximation 
a rescattering phase exists only for positive binding energy.

Consistency of the approach requires that the infrared singularities
in the coefficient functions of the $P$-wave colour-singlet
operators cancel with the $1/\epsilon$ pole in (\ref{octetloopfn}).
In this case we may interpret the singularities and corresponding
$\mu$-dependence as factorization scale dependence that cancels
in the unambiguous sum of the two contributions. Including the
tree-level matching coefficients for the colour-octet operators
(\ref{treematching}) and making use of (\ref{eq:13}), the
cancellation condition reads
\begin{equation}
{\cal C}^A_1(\,^{2S+1}P_J^{(1)}) -
{\cal C}^A_\parallel(\,^{2S+1}S_S^{(1)})
\,\frac{\alpha_s}{4\pi}\,\frac{C_F}{N_c}\,
{\cal D}_{sv}(\,^{2S+1}\!P_J^{(1)})\,F_{sv} = \mbox{finite}.
\end{equation}
Inserting results from (\ref{eq:13}), (\ref{eq:15}), (\ref{octetloopfn}),
one checks that the $1/\epsilon$ terms do indeed cancel.
In particular, the vanishing of ${\cal D}_{sv}(\,^3\!P_1^{(1)})$, hence
the complete absence of a soft vertex contribution, for the
$\chi_{c1}$ case is consistent with the absence of an IR divergence in
the loop coefficient $f[\,^3P_1]$.

\subsection{Soft spectator-scattering}

Next we calculate the spectator-scattering contribution to the colour-octet
matrix elements. We first consider the part of the amplitude shown in
Figure~\ref{fig3}, before attaching the gluon to the spectator quark
line. In the present tree approximation, the gluon momentum $k$ is 
the difference between the momentum of the antiquark in the
kaon, $p_{\bar q}^K$ and the spectator-antiquark momentum $p_{\bar q}^B$ 
in the $B$ meson. All components of these momenta
involve factors of $\Lambda$, except for $n_+\cdot p_{\bar q}^K=\bar y n_+ p
=2 \bar y E_K$. Since $m_c v^2\gg \Lambda$ by assumption, we may drop all
small components (except in the denominator of the gluon propagator, 
which would be exactly zero) and approximate $k$ by
\begin{equation}
k^\mu =n_+\cdot k\,\frac{n_-^\mu}{2}
\label{ksimp}
\end{equation}
with $n_+\cdot k \sim m_c v^2$, since $k$ is soft. This implies that
$\bar y\sim v^2$, so soft-spectator scattering corresponds to an
endpoint configuration, in which very little momentum is transferred
to the spectator antiquark. Almost all of the kaon's momentum is
carried by the quark generated at the $b\to s$ vertex. The gluon
virtuality is given by $k^2=-n_+ \cdot p_{\bar q}^K \,
n_-\cdot p_{\bar q}^B = -2 E_K \bar y \omega$.

The starting expression for the matrix element is
\begin{eqnarray}
\langle H \bar K| \mathcal{O}^A_\perp(\,^{2S+1}S_J^{(8)})|
\bar B \rangle|_{\mbox{{\scriptsize spect}}}
&=&
\sqrt{2 M_H}\,\frac{1}{\sqrt{2 N_c}}
\int \frac{d^3\bfm{\ell}}{(2\pi)^3}\,\sqrt{\frac{3}{4\pi}}\,
\tilde R_{21}(\ell)
\,\frac{1}{N_c}\int_0^\infty \!d\omega \,f_B\,\phi_{B+}(\omega)
\nonumber\\
&& \hspace*{-4cm}
\times \,
\frac{1}{N_c}\int_0^1 dy\,f_K\,\phi_K(y)
\,\frac{(-i) g_{\rho\alpha}}{-2 E_K \bar y \omega}
\,\mbox{tr}\left(\Lambda[H]\,\frac{1+\not\!v}{2}\, A_{c\bar c}(E_H,\ell_\top)
\,\frac{1-\not\!v}{2}\right)
\nonumber\\
&& \hspace*{-4cm}
\times \,(-i) g_s \,\mbox{tr}\,(T^A T^B) \,
\frac{i}{4}\left(-\frac{i}{4}\right) M_B\,
\mbox{tr}\left(\not\!p \,\gamma_5 \gamma^\mu_\perp (1-\gamma_5)\,
\frac{1+\not\!w}{2}\!\not\!n_+\gamma_5\gamma_\alpha\right),
\label{octetspect1}
\end{eqnarray}
where the second trace arises from the gluon coupling to the spectator
quark and the projections on the leading-twist light-cone distribution
amplitudes of the kaon and the $B$ meson.
The attachment of the soft gluon to the $c\bar c$ lines,
$ A_{c\bar c}(E_H,\ell_\top)$, is the
same for the vertex and spectator contribution. Substituting (\ref{ksimp})
into (\ref{ccamp}) shows that the gluon index $\rho$ must be
transverse.
The corresponding
matrix elements of the operators $\mathcal{O}^A_\parallel(^{2S+1}S_J^{(8)})$
have $\!\not\!n_+$ instead of $\gamma_{\perp\mu}$ in the
second trace. The trace then vanishes (also for the projection on
$\phi_{B-}(\omega)$, since $\alpha$ must be transverse),
so at this order
\begin{equation}
\langle H \bar K| \mathcal{O}^A_\parallel(\,^{2S+1}S_J^{(8)})|
\bar B \rangle|_{\mbox{{\scriptsize spect}}} =0.
\end{equation}

Further evaluation of (\ref{octetspect1}) is straightforward: perform
the traces; convert the $\omega$-integral into $1/\lambda_B$ (\ref{eq:24});
do the $\bfm{\ell}$ angular and then the radial integral. To facilitate
the comparison with (\ref{hardspecamp1}), we provide the final result
for the partial amplitude
\begin{equation}
\mathcal{A}^{\mbox{{\scriptsize{soft spectator}}}}_{B\rightarrow
H(^{2S+1}P_J) K} =\frac{G_F}{\sqrt{2}} \,V_{cb}V^*_{cs} \,
{\cal C}_{\perp}^A(\,^{2S+1}S_S^{(8)})\,
\langle H\bar K| \mathcal{O}_{\perp}^A(\,^{2S+1}S_S^{(8)})
|\bar B\rangle|_{\mbox{{\scriptsize spect}}}
\end{equation}
rather than the matrix element itself:
\begin{eqnarray}
\mathcal{A}^{\mbox{{\scriptsize{soft spectator}}}}_
{B\rightarrow H(^{2S+1}P_J) K} &=&
\frac{G_F}{\sqrt{2}} \,V_{cb}V^*_{cs} \,
\frac{\pi \alpha_s C_F}{N_c}\,
\langle H(^{2S+1}P_J)|\mathcal{O}(^{2S+1}P_J^{(1)}) |0\rangle
\nonumber\\
&& \hspace*{-2.6cm}\times\,\frac{f_K f_B M_B}{m_b \lambda_B}
\,\frac{2 C_1}{N_c}\,B[\,^{2S+1}P_J^{(1)}]\,
\int_0^1 d y \,\phi_{\bar K}(y)\,
\left(\sqrt{-\left(\bar y+\frac{2\sqrt{z} E_H}{m_b(1-z)}
\right)}+\frac{\gamma_B}{m_b\sqrt{1-z}}\right)^{\!\!-4}\!\!\!.
\nonumber\\[-0.2cm]
\label{softspecamp1}
\end{eqnarray}
Here $B[\,^{2S+1}P_J^{(1)}]$ is given by the same expression as defined 
in (\ref{eq:23}), (\ref{cb1}). Notice that there is no soft 
spectator-scattering contribution to the $h_c$ final state, as 
 $B[\,^{1}P_1^{(1)}]=0$, which will be important in the numerical 
analysis.

We compare the integral over the kaon 
distribution amplitude to (\ref{hardspecamp1}). While the integrand there 
was applicable to $y$ not near 1 and exhibited a logarithmic endpoint
divergence as $y\to 1$, the present integrand is appropriate
only to $1-y\sim v^2$, i.e. in the endpoint region. There is no divergence
here as $y\to 1$. However, for $\bar y\gg v^2$ the integrand has the
same logarithmic behaviour $\int dy\,\phi_{\bar K}(y)/\bar y^2$ as 
does the hard-spectator contribution for $\bar y\ll 1$.
In (\ref{subtractend}), (\ref{finalconv}) we regulated the endpoint
divergence in hard spectator-scattering by cutting off the
$y$ integral above $1-\mu$. This corresponds to a hard factorization
scale in the energy of the gluon that connects to the spectator
quark. The spectator-scattering contribution to the colour-octet
matrix element originates precisely from the energy region that
was left out in (\ref{finalconv}), thus the correct interpretation
of the $y$-integral in (\ref{softspecamp1}) is
$\int_0^1 dy \to \int_{1-\mu}^1 dy$. To combine with  (\ref{finalconv})
we must evaluate the regularized version of (\ref{softspecamp1})
up to terms of order $v^2/\mu$. This allows us to approximate
$\phi_{\bar K}(y)\approx -\bar y\,\phi_{\bar K}^\prime(1)$,
resulting in
\begin{eqnarray}
&&\int_{1-\mu}^1 d y \,\phi_{\bar K}(y)\,
\left(\sqrt{-\left(\bar y+\frac{2\sqrt{z} E_H}{m_b(1-z)}
\right)}+\frac{\gamma_B}{m_b\sqrt{1-z}}\right)^{\!\!-4}
\nonumber\\
&& \hspace*{1cm}\approx\,- \phi_{\bar K}^\prime(1)
\int_0^{\mu} d y \,
\left(\sqrt{-\left(y+\frac{2\sqrt{z} E_H}{m_b(1-z)}
\right)}+\frac{\gamma_B}{m_b\sqrt{1-z}}\right)^{\!\!-4}
\nonumber\\
&& \hspace*{1cm}
\approx\,\phi_{\bar K}^\prime(1)\left(-\ln\mu + F_{ss}\right),
\end{eqnarray}
where
\begin{equation}
F_{ss} = -\ln(1-z) +1+i \pi +\Delta F,
\label{fss}
\end{equation}
and $\Delta F$ is given in (\ref{octetloopfn}). Comparing
(\ref{hardspecamp1}) to (\ref{softspecamp1}), together with
(\ref{finalconv}), the previous equation demonstrates that the
regulator-dependent $\ln\mu$ terms cancel. We may therefore conclude
that the
endpoint singularity is hard spectator-scattering does not
violate factorization, since it can be factorized into the
colour-octet matrix elements.

\subsection{Further remarks on the endpoint
singularity}

A rigorous understanding of endpoint singularities in convolution
integrals would enhance the predictivity of QCD factorization approaches 
for exclusive $B$ decays considerably. It would also provide meaning to formal
factorization ``theorems'' derived in soft-collinear effective theory,
which generically result in ill-defined convolutions, with exceptions
in many leading-power applications. Despite several attempts
\cite{Beneke:2003pa,Becher:2003qh,Manohar:2006nz} there is currently
no satisfactory framework for factorizing endpoint divergences and
for associating them with well-defined operator matrix elements.

The calculation of the $B\to \chi_{cJ} K$ decay amplitudes presented
in this paper provides the first example, where an endpoint singularity
in a hard-scattering convolution integral can be factored consistently
(at least at the leading order) into precisely defined objects,
the colour-octet matrix elements. The example does not quite represent
what is required for other cases, such as the $B\to\pi$ form factor,
since in $B\to \chi_{cJ} K$ the endpoint divergence arises from
factorization at the scale $m_c v^2$, not $\Lambda$.\footnote{This 
aspect is similar to the discussion of the $B\to \eta_c$ form factor 
in \cite{Bell:2005gw}, which also exhibits a calculable endpoint 
logarithm.}  Nevertheless,
it is worthwhile to collect some observations on the structure of
the endpoint contribution.

\begin{itemize}
\item The endpoint contribution is proportional to
$\phi_{\bar K}^\prime(1)$, the derivative of the distribution amplitude 
at the endpoint.\footnote{As in the expression 
for the $B\to \pi$ form factor in the heavy quark limit of its 
light-cone QCD sum rule representation \cite{Bagan:1997bp}.} 
This is because the endpoint region is of size $v^2$ rather than
$\Lambda/m_b$, hence it is justified to describe the quarks
in the kaon by collinear quark fields. In general, we do not
expect the distribution amplitudes to be relevant in the
endpoint region, since one of the quarks does not carry a
collinear momentum.
\item The factor $1/\bar y$ that renders the hard-spectator
convolution integral divergent, originates from the expansion of
the charm propagators. It is therefore not possible to associate
the endpoint contribution with a matrix element involving only
the kaon state, as in the case of the light-cone distribution
amplitude. Rather it reflects a large non-factorizing contribution
to the entire process from the scale $m_c v^2$.
\item  The endpoint contribution contains a large rescattering
phase as seen from (\ref{fss}). This observation casts doubt
on the correctness of a claim made in \cite{Arnesen:2006vb}
that the power-suppressed weak annihilation contributions to
charmless decays are real in first approximation. In fact, applying
the prescription of endpoint subtraction used in \cite{Arnesen:2006vb}
to $B\to \chi_{cJ} K$ would simply set the soft-spectator
scattering contribution to zero. This implies an uncanceled
subtraction scale dependence proportional to $\ln \mu$, which
is also present in the result of \cite{Arnesen:2006vb}. More
important to the claim,
it would miss the soft spectator-rescattering phase.
\end{itemize}

\section{Estimates of branching ratios}
\label{pheno}

Before discussing the numerical results, we present an expression for
the sum of the different contributions to the amplitude:
\begin{equation}
\mathcal{A}_{B\rightarrow H(^{2S+1}P_J) K} =
\frac{G_F}{\sqrt{2}} \,V_{cb}V^*_{cs} \,
\langle H(^{2S+1}P_J)|\mathcal{O}(^{2S+1}P_J^{(1)}) |0\rangle
\Big[A_V+A_S\Big],
\end{equation}
where separating the vertex corrections from the spectator-scattering
corrections we have
\begin{eqnarray}
A_V  &=& 2 M_B \left(1-\frac{M_H^2}
{M_B^2}\right) f_+^{BK}(M_H^2)\,\Bigg[
\left({\mathcal{C}}_0^A(^{2S+1}\!P_J^{(1)}) +
{\mathcal{C}}_1^A(^{2S+1}\!P_J^{(1)})\right)
\nonumber\\
 && -
\,\frac{\alpha_s}{4\pi}\,\frac{C_F}{N_c}\,
{\cal D}_{sv}(\,^{2S+1}\!P_J^{(1)})\,F_{sv}\,
{\cal C}^A_\parallel(\,^{2S+1}S_S^{(1)})\Bigg],
\nonumber\\
A_S &=&  \frac{\pi \alpha_s C_F}{N_c}\,
\frac{f_K f_B M_B}{m_b \lambda_B}\,\frac{2 C_1}{N_c} \,
\Bigg\{
C^B[\,^{2S+1}\!P_J\,]
\int_0^{1} d y \,\frac{\phi_{\bar K}(y)}{\bar y}
\nonumber\\
&& \hspace*{0cm}+ \, B[\,^{2S+1}\!P_J\,] \left(
\phi_{\bar K}^\prime(1) F_{ss}+
\int_0^1 dy\,\frac{\phi_{\bar K}(y)+\bar y \,
\phi_{\bar K}^\prime(1)}{\bar y^2} \right)
\Bigg\},
\label{fullamp}
\end{eqnarray}
and all quantities have been defined previously. The branching fraction,
extending the leading-order expression (\ref{brLO}), is given by
\begin{eqnarray}
\mbox{Br}\,(\bar B\rightarrow H(^{2S+1}P_J) \bar K)
&=& \tau_{B}\,\frac{G_F^2}{2}\,|V_{cb} V_{cs}^*|^2\,
\frac{M_B^2-M_H^2}{16\pi M_B^3}\,\mbox{PolSum}[H(^{2S+1}P_J)]\,
\nonumber\\
&&\times\,
\Big[\langle\mathcal{O}(^{3}P_0^{(1)})\rangle\Big]^2
\,\Big|A_V+A_S\Big|^2.
\label{brNLO}
\end{eqnarray}

\subsection{Parameters}

The numerical result depends on the couplings $G_F=1.16639 \cdot 10^{-5}\,
\mbox{GeV}^{-2}$, $|V_{cb} V_{cs}| = 40.4 \cdot 10^{-3}$. The QCD scale
is $\Lambda_{\rm \overline{MS}}^{(5)}=225\,$MeV ($\overline{\rm MS}$
scheme, five quark flavours) and next-to-leading logarithmic running
of the strong coupling and Wilson coefficients is used. At $\mu=3\,$GeV:
$\alpha_s(3\,\mbox{GeV}) = 0.2503$, $C_1(3\,\mbox{GeV})=1.105$,
$C_2(3\,\mbox{GeV}) = -0.2366$. The renormalization scale for these
quantities is denoted by $\mu$. However, in the strong coupling
that multiplies the spectator scattering
term we use the intermediate scale $\mu_h=\sqrt{0.5\,\mbox{GeV} \,\mu}$,
and in the expression for the inverse Bohr radius (\ref{bohrradius})
we imagine choosing the scale of $\alpha_s$ (or $m_c$)
such that $\gamma_B = 500\, \mbox{MeV}.$ The values of the quark masses
$m_b$, $m_c$ will be discussed below.

The meson masses are $M_B=5.279\,$GeV,
$M_{\chi_{c0}}=3.415\,$GeV,
$M_{\chi_{c1}}=3.511\,$GeV,
$M_{\chi_{c2}}=3.556\,$GeV,
$M_{h_{c}}=3.525\,$GeV. The derivative of the wave function at the origin
can be determined from $\chi_{cJ}$ decays and takes the value
$|R_{21}^\prime(0)|^2 = 0.050\,\mbox{GeV}^5$
\cite{Beneke:1998ks,Brambilla:2004wf}. The kaon and $B$-meson decay
constants are $f_K=160\,$MeV and $f_B=210\,$MeV, respectively, the
moment of the $B$-meson distribution amplitude is assumed to
take the small value $\lambda_B=200\,$MeV that is favoured
by the large rates of colour-suppressed charmless $B$ 
decays~\cite{Beneke:2005vv}. The
$B\to K$ form factor is parameterized following~\cite{Ball:2004ye}
in the form
\begin{equation}
f_+^{BK}(q^2) = \frac{0.162}{1-q^2/M_{B_s}^2}+
\frac{0.173}{(1-q^2/M_{B_s}^2)^2}
\qquad (M_{B_s}=5.41\,\mbox{GeV}).
\label{d23}
\end{equation}
Many of these parameters have significant theoretical errors, but in view
of other uncertainties discussed below, they are less relevant, except
for the parameter $\lambda_B$, where values twice as large are not ruled
out theoretically.
Finally, we expand the kaon light-cone distribution amplitude
into Gegenbauer polynomials
\begin{equation}
\phi_{\bar K}(x)\,=\, 6 x\bar x \left[1+\sum\limits_{n=1}^{\infty}
a_n^{\bar K} C^{(3/2)}_n(2 x-1)\right]\,,
\end{equation}
and truncate the expansion at order $n=2$. The first two Gegenbauer moments
are $a_1^{\bar K}(2\,\mbox{GeV})=0.06\pm 0.06$ and
$a_2^{\bar K}(2\,\mbox{GeV})=0.2\pm 0.15$ \cite{Khodjamirian:2004ga,Braun:2004vf,Ball:2005vx,Ball:2006wn,Braun:2006dg,Boyle:2006pw},
while using ``asymptotic''
distribution amplitudes amounts to setting the Gegenbauer moments
to zero. In terms of Gegenbauer moments the expressions appearing
in (\ref{fullamp}) read
\begin{eqnarray}
&& \phi_{\bar K}^\prime(1) =
(-6) \,\left(1 + 3 a_1^{\bar K} + 6 a_2^{\bar K}\right),
\nonumber\\
&&I_1\equiv \int_0^1 dy \,\frac{\phi_M(y)}{\bar y}
= 3\,\left(1 + a_1^{\bar K} + a_2^{\bar K}\right),
\nonumber\\
&& I_2\equiv \int_0^1 dy\,\frac{\phi_{\bar K}(y)+\bar y \,
\phi_{\bar K}^\prime(1)}{\bar y^2}  =
(-6) \,\left(1 + 6 a_1^{\bar K} + 16 a_2^{\bar K}\right).
\end{eqnarray}
While $I_1$ is well-behaved, $ \phi_{\bar K}^\prime(1)$
and $I_2$ exhibit a very large sensitivity to the higher Gegenbauer
moments. Comparing the maximal value of $I_2$ to its asymptotic
one, we find $I_2^{\rm max}/I_2^{\rm as} = 7.32$ and it is not
clear whether the Gegenbauer expansion is converging at all. A
consequence of this is that the size of the spectator-scattering
amplitude is uncertain by a factor of several (including the
uncertainty in $\lambda_B$) for
$\chi_{cJ}$, where $B[\,^{2S+1}\!P_J\,]$ is not vanishing.

\subsection{Results}

Given the large ambiguities mentioned above, but also the fact that
our calculation relies on the unrealistic assumption that charmonium
is a Coulomb bound state, we do not expect reliable quantitative
results for the $B\to (\chi_{cJ},h_c) K$ branching fractions. Instead
we address the questions
\begin{itemize}
\item[1)] Are large corrections to naive factorization expected
theoretically?
\item[2)] Why are the $\chi_{c2}K$ and $h_c K$ final states
suppressed relative to $\chi_{c0}K$, $\chi_{c1}K$?
\end{itemize}
which are of interest given observations summarized
Table~\ref{tab1}. Our calculation results in
exactly the same decay rates for charged and neutral $B$ decay.
Thus, branching fractions of pairs of related decays differ 
only by the  lifetime ratio
$\tau_{B_u}/\tau_{B_d}$. In the following we consider only $B_d^0$ decay
using $\tau_{B_d}=1.53 \cdot 10^{-12}\,\mbox{s}$.

\begin{figure}[t]
\begin{center}
\hspace*{-0.5cm}
\includegraphics[width=0.49 \linewidth]{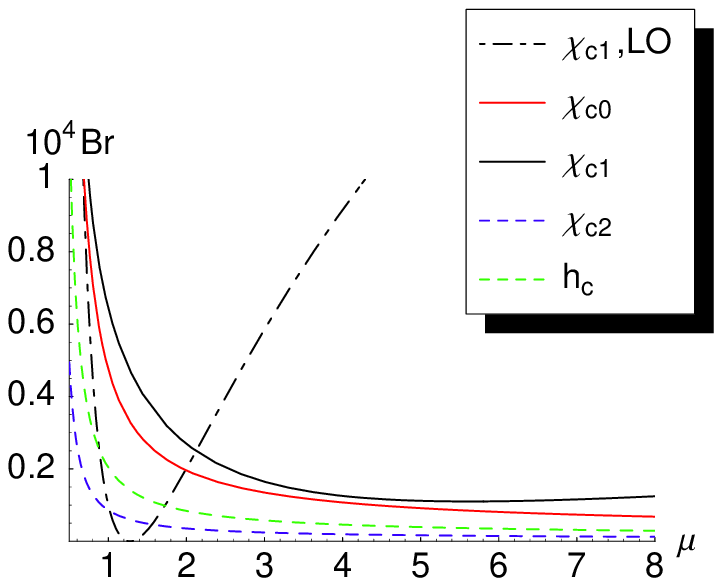}
\includegraphics[width=0.49 \linewidth]{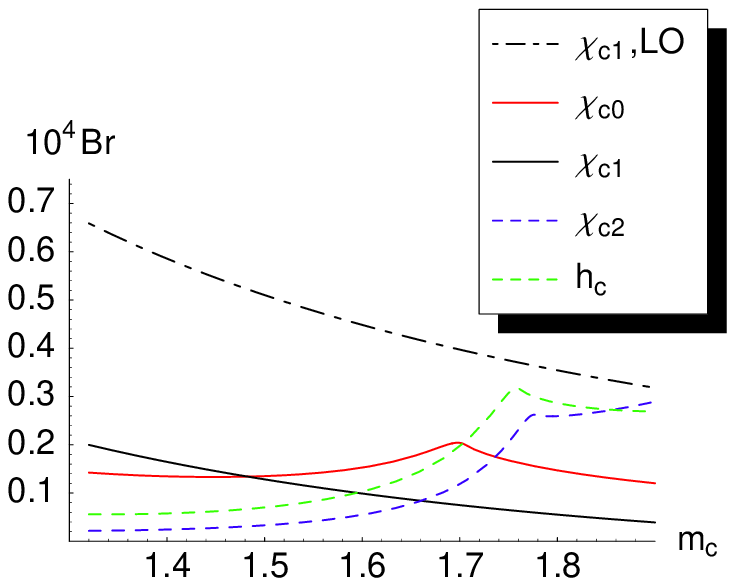}
\caption{Branching fractions in units of $10^{-4}$, when the 
spectator-scattering contribution is set to zero: renormalization 
scale (left) and charm quark mass (right) dependence.}
\label{fig:plots1}
\end{center}
\end{figure}

We begin by discussing the dependence of the branching fractions on
the various inputs, when we neglect the spectator-scattering
term entirely. The scale-dependence, adopting the quark masses values 
$m_b=4.8\,$GeV, $m_c=1.4\,$GeV, is shown in the left plot of
Figure~\ref{fig:plots1}. There is a large cancellation between
the tree level and one-loop contribution to the coefficient function
relevant to the $\chi_{c1}$ final state resulting in a very small
branching fraction. The branching fractions for the other final states 
are also quite small, not exceeding a few times $10^{-5}$ with 
$h_c$ and $\chi_{c2}$ being even smaller than the other two. 
At this point we can already conclude that corrections to naive 
factorization are order one effects, providing a positive answer to 
the first question above, as is in fact expected for colour-suppressed
decay modes. The final-state dependence might be similar to the data, 
but this cannot be the complete story, since the $\chi_{c0}$ and 
$\chi_{c1}$ branching fractions fall short of the data by about 
an order of magnitude.

The scale dependence of the NLO result remains significant, simply
because the LO term for $\chi_{c1}$ is canceled, and
there is no LO term for the other final states. The scale dependence
is exactly of the form $(\alpha_s C_1)^2$ for $\chi_{c0}$, $\chi_{c2}$,
and $h_c$ and approximately so for  $\chi_{c1}$. This causes
an uncertainty of a factor of 2 when $\mu$ is varied between
$2\,$GeV and $5\,$GeV, and larger if one allows smaller scales. However,
below about $1.5\,$GeV the scale-dependence blows up as seen
in the Figure. In the following we fix $\mu=3\,$GeV. Results for
other choices of $\mu$ can be obtained approximately by multiplying
with $(\alpha_s C_1)^2(\mu)/(\alpha_s C_1)^2(3\,\mbox{GeV})$. (This 
remains true, when the spectator-scattering terms are added back.)
Other significant parameter dependences arise from the quark mass
values. The dependence on $m_c$ is more important than the one on
$m_b$, so we fix $m_b=4.8\,$GeV in the following. The charm quark 
mass dependence of the branching fractions, still omitting spectator
scattering, is shown in the right plot of Figure~\ref{fig:plots1}, 
from which it is seen that the size of the $\chi_{c0}$, $\chi_{c1}$ 
branching fractions versus $\chi_{c2}$, $h_c$ reverses as $m_c$ 
increases. The charm quark mass here is the pole mass, which is a 
poorly defined quantity in perturbation theory, due to large 
radiative corrections.
Typical values are $m_c = 1.4 \ldots 1.7\,$GeV. There may be good
reason to choose larger values here, since the charm quark pole
mass controls the binding energy $E_H=M_H-2 m_c$, which should be
negative in the approximation of charmonium as a non-relativistic
bound state, and is negative in reality when measured relative
to the $D \bar D$ threshold. From the Figure it is evident that 
with the NLO vertex correction alone it is not possible to explain 
the experimental data, since the  $\chi_{c0}$, $\chi_{c1}$ branching 
fractions are too small, even allowing for theoretical uncertainties 
in the $B\to K$ form factor or the charmonium  wave function. 

\begin{figure}[p]
\begin{center}
\includegraphics[width=0.49 \linewidth]{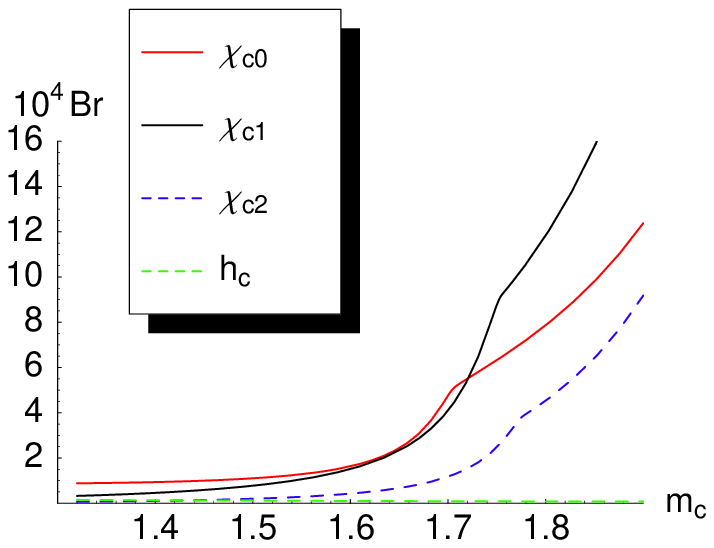}
\includegraphics[width=0.49 \linewidth]{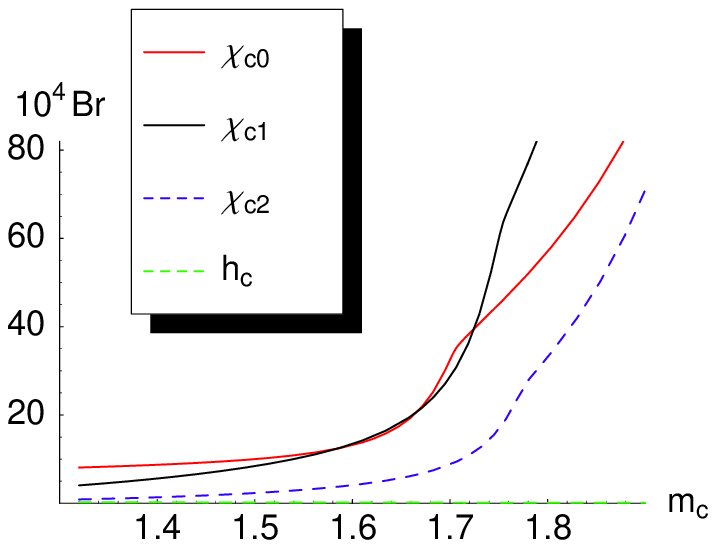}
\caption{Charm quark mass dependence of the branching fractions 
in units of $10^{-4}$ from spectator scattering 
only. Left for asymptotic kaon distribution amplitude 
($a_1^{\bar K}=a_2^{\bar K}=0$), right for $a_1^{\bar K}=0.06$ and 
$a_2^{\bar K}=0.2$.}
\label{fig:plots2}
\end{center}
\end{figure}

\begin{figure}[p]
\begin{center}
\includegraphics[width=0.7 \linewidth]{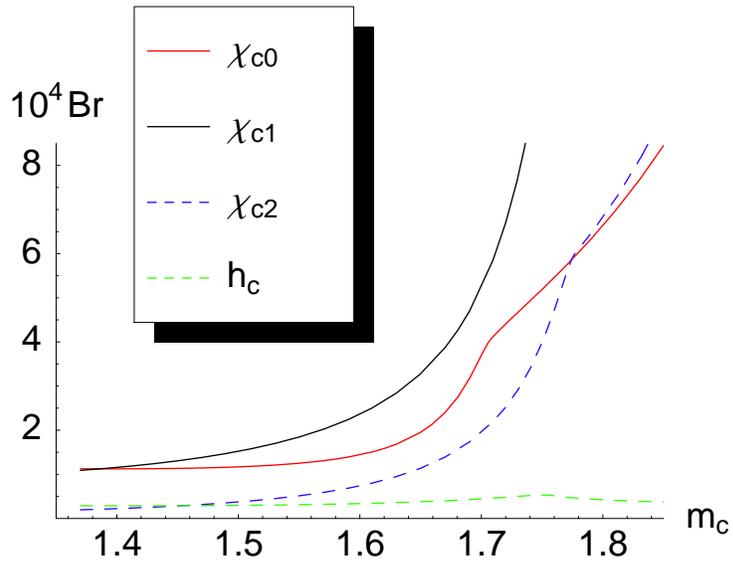}
\caption{Charm quark mass dependence of the branching fractions 
in units of $10^{-4}$ including all contributions.}
\label{fig:plots3}
\end{center}
\end{figure}

Next we imagine that the branching
fractions are given by the spectator-scattering term alone. The largest
parameter dependences now arise from the charm quark mass, the Gegenbauer 
moments of the kaon light-cone distribution amplitude, and the 
$B$ meson distribution amplitude parameter $\lambda_B$. 
In Figure~\ref{fig:plots2} we show the
$m_c$ dependence for asymptotic distribution amplitudes (left) and
for $a_1^{\bar K}=0.06$, $a_2^{\bar K}=0.2$ (right). The branching fractions
grow rapidly with  $m_c$, when the spectator amplitude becomes
dominated by the imaginary part from the colour-octet contributions.
We also observe a huge dependence on the Gegenbauer moments, confirming
the expectation that the expansion may be invalid. Even 
$a_2^{\bar K}=0.2$ leads to unacceptably large branching fractions. 
These effects are less pronounced when $\lambda_B$ is larger, since 
the spectator-scattering branching fraction shown in Figure~\ref{fig:plots2} 
is proportional to $1/\lambda_B^2$. 
Independent of these uncertainties, we always find that
spectator-scattering is a small effect for $h_c$, and larger for
$\chi_{c0}$, $\chi_{c1}$ than for the other two final states.

\begin{table}[t]
\begin{center}
\begin{tabular}{c|c|c|c|c} \hline
&&&&\\[-0.4cm]
$m_c/\mbox{GeV}$ &
$\chi_{c0}K^0 $ & $\chi_{c1}K^0 $ & $\chi_{c2}K^0 $ & $h_cK^0$
\\[0.2cm]
\hline
&&&&\\[-0.4cm]
1.45 &
$1.13$  &
$1.31$  &
$0.28$  &
$0.29$
\\[0.2cm]
1.50 &
$1.17$  &
$1.52$  &
$0.37$  &
$0.30$
\\[0.2cm]
1.55 &
$1.25$  &
$1.85$  &
$0.51$  &
$0.31$
\\[0.2cm]
1.60 &
$1.44$  &
$2.36$  &
$0.74$  &
$0.34$
\\[0.2cm]
1.65 &
$1.95$  &
$3.28$  &
$1.14$  &
$0.38$
\\[0.2cm]
1.70 &
$3.70$  &
$5.23$  &
$1.95$  &
$0.44$
\\[0.2cm]
1.75 &
$5.19$  &
$10.31$  &
$3.98$  &
$0.53$
\\[0.2cm]
\hline
\end{tabular}
\caption{$B^0_d\to H K^0$ branching fractions in units of $10^{-4}$ for
various charm quark mass inputs. Asymptotic distributions amplitudes,
$\lambda_B=200\,$MeV and $m_b=4.8\,$ are fixed.
\label{tab2}}
\end{center}
\end{table}

When we now add both contributions together, including the interference
term, we obtain the result shown in Figure~\ref{fig:plots3} for
asymptotic distribution amplitudes. For
$m_c$ in the range from $1.5\,$GeV to $1.7\,$GeV, this suggests the 
interpretation that the $\chi_{c0}$, $\chi_{c1}$ final states are 
dominated by spectator scattering, more precisely by the 
spectator-scattering contribution to the colour-octet matrix element.
The smallness of the $h_c$ branching fraction is explained by the 
absence of such a contribution (at leading order) for this final state. 
The $\chi_{c2}$ case is intermediate with a rapidly rising branching fraction 
in the interesting charm-quark mass window. Numerical results 
for some values of $m_c$ are provided in Table~\ref{tab2}. We
emphasize that in addition to the charm-quark mass dependence displayed
explicitly there are further large theoretical uncertainties related
to scale-dependence, which shifts all branching fractions uniformly
as described above, to $m_b$-dependence, to the Gegenbauer moments, and 
to $\lambda_B$. There are some parameter degeneracies that allow 
making $a_2^{\bar K}$ and $\lambda_B$ simultaneously larger. In view 
of these uncertainties the main conclusion of the numerical analysis is
that there are reasonable regions of parameter space ($m_c 
\approx 1.65\,$GeV, small $\lambda_B$, and asymptotic 
kaon distribution amplitude), where the theoretical calculations in 
our model for the colour-octet matrix elements are in qualitative 
agreement with the experimental data, namely the existence of large 
contributions beyond naive factorization, and the suppression of the
$\chi_{c2}K$ and $h_c K$ modes. From Table~\ref{tab2} we conclude 
that the small $h_c$ branching fraction 
is a robust feature of our results, but we find it difficult 
to explain the strong $\chi_{c2}$ suppression seen in the data compiled in 
Table~\ref{tab1}, while maintaining the sizeable 
$\chi_{c1}$ branching fraction.

\section{Conclusion}
\label{sec4}

We revisited exclusive $B$ decays to $P$-wave charmonia motivated
by previous studies
\cite{Song:2002mh,Song:2003yc,Pham:2005ih,Meng:2005fc,Meng:2005er,Meng:2006mi}
of these decays in the QCD factorization framework that reported
a violation of factorization. In contrast, we find that after
accounting for colour-octet operators, which, contrary to the case
of charmless decays, are not suppressed
by $\Lambda/m_b$ due to the existence of the charmonium binding energy
scale, QCD factorization is recovered, at least at order $\alpha_s$.
The infrared divergences found in previous calculations can be
subtracted consistently into the matrix elements of these operators.
This includes the endpoint divergence that is found in the
unsubtracted coefficient function associated with
spectator-scattering. Our calculations demonstrate that the
endpoint contribution, now contained in the colour-octet
matrix element, can lead to a large rescattering phase. These observations 
may be of conceptual interest, since it is presently still unclear 
in the general case, whether and how endpoint divergences that often 
appear in convolutions in collinear factorization formulas can be 
absorbed into well-defined non-perturbative objects and what these 
objects are. We find it plausible that factorization of $B$ decays 
to $P$-wave charmonium extends to 
higher orders in the coupling expansion when $m_c v^2\gg \Lambda$,  
in view of the argument presented in \cite{Beneke:2000ry}; nonetheless, 
it would be of great interest to verify the factorization of endpoint 
divergences beyond the tree-approximation to the hard-scattering 
sub-graph.  

Previous numerical estimates of the branching fractions to $P$-waves
relied on ad hoc treatments of the infrared regulator dependence.
In the present framework, this is unnecessary, but an estimate of
the colour-octet operator matrix elements is needed, which may even
be the largest contribution to the decay amplitude. To this end we
adopted a description of charmonium as a Coulomb bound state, which
corresponds to the formal heavy quark limit. In practice, this limit
is probably unreliable, and our results do indeed exhibit large
theoretical uncertainties. Nevertheless, we find that for plausible
theoretical inputs it is possible to reproduce qualitatively what we 
consider to be the most interesting features of current experimental data:
suppression of the $\chi_{c2}K$ and $h_c K$ final states and
amplitudes that must be dominated by terms beyond naive factorization, 
though the suppression of  $\chi_{c2}K$ is not as strong as observed. 
An interesting avenue to pursue in the future might be to 
consider the colour-octet matrix elements as unknown non-perturbative
parameters, which is more realistic in view of $m_c v^2\sim \Lambda$,
and to exploit the constraints imposed by spin-symmetry on the leading 
contributions to these matrix elements.

\vspace*{1em}

\noindent
\subsubsection*{Acknowledgement}
This work is supported in part by the 
DFG Sonder\-forschungsbereich/Transregio~9 
``Computergest\"utzte Theoretische Teilchenphysik'' 
and the Swiss National Science Foundation (SNF). 
M.B. acknowledges hospitality from the University of Z\"urich and the 
CERN theory group, where part of this work was performed.

%\bibliography{paper}

\end{document}